\documentclass[twocolumn,amsmath,amssymb,floatfix,prb,shownopacs,footinbib,superscriptaddress]{revtex4}
\usepackage{amsmath,amssymb,natbib,bm,graphicx,url,epsf,color}
\usepackage[british]{babel}
\usepackage{wasysym}
\usepackage{MnSymbol}

\newcommand{\Ginf}{G_{\infty}}

\newcommand{\Gmax}{G_\mathrm{max}}

\begin{document}

\title{Tomonaga-Luttinger physics in electronic quantum circuits}

\author{S. Jezouin}
\affiliation{CNRS / Univ Paris Diderot (Sorbonne Paris Cit\'e), Laboratoire de Photonique et de Nanostructures
(LPN), route de Nozay, 91460 Marcoussis, France}
\author{M. Albert}
\affiliation{CNRS / Univ Paris Sud, Laboratoire de Physique des Solides
(LPS), 91405 Orsay, France}
\author{F.D. Parmentier}
\affiliation{CNRS / Univ Paris Diderot (Sorbonne Paris Cit\'e), Laboratoire de Photonique et de Nanostructures
(LPN), route de Nozay, 91460 Marcoussis, France}
\author{A. Anthore}
\affiliation{CNRS / Univ Paris Diderot (Sorbonne Paris Cit\'e), Laboratoire de Photonique et de Nanostructures
(LPN), route de Nozay, 91460 Marcoussis, France}
\author{U. Gennser}
\affiliation{CNRS / Univ Paris Diderot (Sorbonne Paris Cit\'e), Laboratoire de Photonique et de Nanostructures
(LPN), route de Nozay, 91460 Marcoussis, France}
\author{A. Cavanna}
\affiliation{CNRS / Univ Paris Diderot (Sorbonne Paris Cit\'e), Laboratoire de Photonique et de Nanostructures
(LPN), route de Nozay, 91460 Marcoussis, France}
\author{I. Safi}
\affiliation{CNRS / Univ Paris Sud, Laboratoire de Physique des Solides
(LPS), 91405 Orsay, France}
\author{F. Pierre}
\email[e-mail: ]{frederic.pierre@lpn.cnrs.fr}
\affiliation{CNRS / Univ Paris Diderot (Sorbonne Paris Cit\'e), Laboratoire de Photonique et de Nanostructures
(LPN), route de Nozay, 91460 Marcoussis, France}

\date{\today}

\maketitle

{\sffamily
In one-dimensional conductors, interactions result in correlated electronic systems. At low energy, a hallmark signature of the so-called Tomonaga-Luttinger liquids (TLL) is the universal conductance curve predicted in presence of an impurity.
A seemingly different topic is the quantum laws of electricity, when distinct quantum conductors are assembled in a circuit. In particular, the conductances are suppressed at low energy, a phenomenon called dynamical Coulomb blockade (DCB).
Here we investigate the conductance of mesoscopic circuits constituted by a short single-channel quantum conductor in series with a resistance, and demonstrate a proposed link to TLL physics.
We reformulate and establish experimentally a recently derived phenomenological expression for the conductance using a wide range of circuits, including carbon nanotube data obtained elsewhere. By confronting both conductance data and phenomenological expression with the universal TLL curve, we demonstrate experimentally the predicted mapping between DCB and the transport across a TLL with an impurity.
}

Despite a very large number of strongly interacting electrons, the low-lying electronic excitations in conventional bulk metals can be described as weakly interacting Fermi quasiparticles, as attested by the remarkable success of Landau's Fermi liquid theory \cite{pinesnozieres1966tql}. This picture breaks down in one-dimensional (1D) conductors, where interactions result in cooperative behaviors \cite{giamarchi2003qp1d}.
According to the TLL theory \cite{tomonaga1950,luttinger1963,mattis1965,haldane1981,giamarchi2003qp1d}, the low-energy elementary excitations in 1D are collective plasmon modes, markedly different from their constitutive individual electrons. This gives rise to intriguing phenomena such as the separation of spin and charge degrees of freedom into distinct elementary excitations propagating at different velocities \cite{giamarchi2003qp1d}; or the charge fractionalization of an injected electron \cite{safi1995}.
Experimentally, indications of TLL physics were observed in such 1D systems as nanotubes \cite{bockrath1999,postma2000}, quantum wires \cite{auslaender2005,jompol2009}, and chains of spins \cite{lake2005,kim2006} or atoms \cite{blumenstein2011}. In addition, this physics is applicable to other many-body phenomena including the fractional quantum Hall effect \cite{wen1992tesfqh,milliken1996,chang2003cll}, the quantum noise in 1D Bose condensates \cite{hofferberth2008}, and the DCB \cite{safi2004ohmic}.

In the present work, we investigate experimentally the DCB conductance suppression across a quantum coherent conductor inserted in a dissipative circuit. This quantum electrodynamics phenomenon, also called zero-bias anomaly, is remarkably similar to a hallmark signature of the collective TLL physics, namely the low-energy conductance suppression in the presence of an impurity (see Ref.~\onlinecite{fisherglazman1997} for a review of TLL more focused on quantum transport). In both situations, the conductance suppression originates from the granularity of charge transfers across the quantum conductor (DCB) or the impurity (TLL). Due to Coulomb interactions, this granularity results in the possible excitation of collective electrical degrees of freedom, which impedes the charge transfers at low energy and therefore reduces the conductance. These collective degrees of freedom are the electromagnetic modes of the surrounding electrical circuit for the DCB, and the plasmon modes for a TLL. In fact, it was shown \cite{safi2004ohmic} that the transport across a short single-channel quantum conductor in series with a pure resistance $R$ can be mapped rigorously onto the transport across a TLL with an impurity. Such circuits therefore provide powerful test-beds for the transport across TLL systems, with many adjustable parameters including the crucial Luttinger interaction coefficient \cite{giamarchi2003qp1d}, given by $K=1/(1+Re^2/h)$. Inversely, as detailed below, the mapping toward a TLL extends the theoretical understanding of DCB.

In order to understand the quantum laws governing electrical transport in mesoscopic circuits composed of distinct quantum components, it is imperative to address the general case of the DCB for arbitrary quantum conductors. This problem remains poorly understood except in the important limit of low transmission coherent conductors realized by tunnel junctions, which can be handled in the theory as a small perturbation to the circuit. For this class of coherent conductors embedded in a linear circuit, extensive experimental and theoretical studies have led to a good understanding \cite{panyukov1988,odintsov1988,nazarov1989,devoret1990dcb,girvin1990dcb,cleland1992dcb,holst1994dcb,joyez1998dcb,pierre2001dcb,hofheinz2011dcb} (see Ref.~\onlinecite{ingold1992sct} for a pedagogical review of the theory).
The tunnel limit was first overcome for relatively small conductance suppressions and low impedance environments compared to the resistance quantum $R_\mathrm{q}=h/e^2\simeq 25.8~\mathrm{k}\Omega$. The striking prediction \cite{yeyati2001dcb,golubev2001dcb} and observation \cite{altimiras2007dcb} are that the conductance suppression is directly proportional to the amplitude of quantum shot noise.
There has also been important progress in the understanding of the regime of relatively strong conductance suppression, where the deviation to classical impedance composition laws are large (see e.g. Refs~\onlinecite{matveev1993tunnel1d,flensberg1993,nazarov1999cbwotj,kindermann2003fcs,safi2004ohmic,golubev2005dcb,zamoum2012}).
In particular, the mapping of DCB to a TLL \cite{safi2004ohmic} opens access to the strong DCB regime for an arbitrary short single-channel quantum conductor in series with a pure resistance of arbitrary value (see also Refs~\onlinecite{lehur2005,florens2007} beyond the short conductor limit). Experimentally, the strong DCB regime was recently explored for circuit impedances comparable to $R_\mathrm{q}$ ($R=13~\mathrm{k}\Omega$ and $26~\mathrm{k}\Omega$) \cite{parmentier2011}; and a generalized phenomenological expression for the transmission of an arbitrary short single-channel quantum conductor embedded in a linear circuit was derived from the data (Eq.~1 in Ref.~\onlinecite{parmentier2011}, see also Eq.~\ref{EqGeneral} in the present article).

Does this phenomenological expression have a deeper significance?  Here we show that the answer is affirmative.
Firstly, we recast Eq.~1 in Ref.~\onlinecite{parmentier2011} as a powerful phenomenological scaling law (Eq.~2) and demonstrate experimentally that it applies to a very wide range of surrounding circuits and single-channel quantum conductors. It is shown to capture the DCB data obtained for series resistances ranging from $R=6~\mathrm{k}\Omega$ to $80~\mathrm{k}\Omega$, and for different realizations of R (using both on-chip chromium resistance or fully transmitted quantum channels).
It is also shown to apply to a very different realization of the single-channel quantum conductor from the quantum point contacts in a Ga(Al)As two-dimensional electron gas measured here and in Ref.~\onlinecite{parmentier2011}. We demonstrate that this phenomenological expression reproduces quantitatively, essentially without fit parameters, the measurements of Finkelstein and coworkers on a carbon nanotube resonant level embedded in a dissipative circuit \cite{mebrahtu2012,mebrahtu2012b}.
Secondly, the origin of the phenomenological expression can be traced back to the TLL collective physics. We establish this link by confronting the full universal conductance scaling curve predicted for a TLL in presence of an impurity \cite{fendley1995,fendley1995b} to the corresponding phenomenological scaling law for a pure series resistance $R$. The agreement is exact at $R = R_\mathrm{q}$, $R\ll R_\mathrm{q}$ and, for arbitrary values of $R$, in the limit of a small single-channel transmission $G R_\mathrm{q}\ll 1$. At intermediate values $R\neq R_\mathrm{q}$, although relatively small deviations emerge, the proposed phenomenological scaling law is found to provide a good approximate expression for the conductance.

Remarkably, the predicted mapping DCB-TLL, here extended theoretically to realistic situations in the presence of a high-frequency (e.g. capacitive) cutoff, is further established by a direct comparison with the DCB conductance data. We demonstrate, with $R=R_\mathrm{q}/4$, a strikingly close agreement, over a broad range of conductances, with the corresponding TLL universal scaling curve computed from the exact thermodynamic Bethe ansatz solution \cite{fendley1995,fendley1995b}.

\begin{figure}[!htbp]
\renewcommand{\figurename}{\textbf{Figure}}
\renewcommand{\thefigure}{\textbf{\arabic{figure}}}
\centering\includegraphics[width=1\columnwidth]{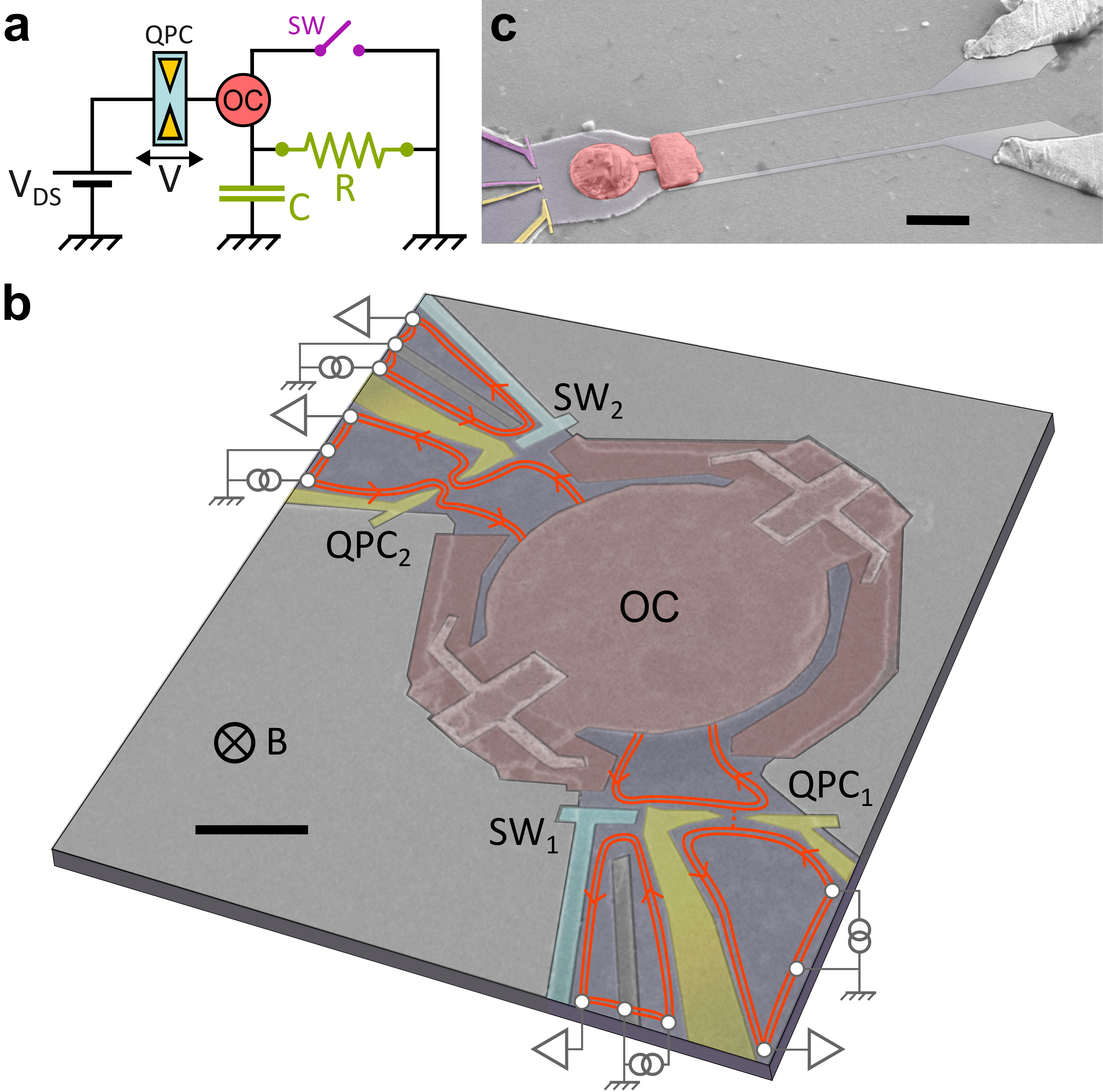}
\caption{\footnotesize
\textbf{Single electronic channel in a resistive environment.} \textbf{(a)}, Schematic circuit for the measured samples. The QPC emulates any single-channel short quantum conductor, $R$ is the on-chip series resistance and $C$ is the geometrical shunt capacitance. Note that $V$ and $V_{\mathrm{DS}}$ are, respectively, the DC voltages across the QPC and the whole circuit. \textbf{(b)}, Colorized SEM micrograph of the sample with a series resistance $R_\mathrm{q}/n$ realized by a second QPC fully transmitting $n$ channels. The 2D electron gas (blue) is separated in two zones by the micron-sized ohmic contact (OC). Each zone comprises a QPC (yellow split-gates) and a short-circuit switch (blue gate) that allows to divert to ground the quantum Hall edge channels (red lines, here at filling factor 2) returning from the central ohmic contact. The black horizontal scale bar is $1~\mu$m long. \textbf{(c)}, Colorized SEM micrograph of the sample with a series resistance $R=6.3~\mathrm{k}\Omega$ realized by two parallel thin chromium wires visible at the right of the ohmic contact (a similar implementation was used for $R=80~\mathrm{k}\Omega$). The black horizontal scale bar is $2~\mu$m long.
\normalsize
}
\label{fig-sample}
\end{figure}

\vspace{\baselineskip}
{\large\noindent\textbf{Results}}\\
{\noindent\textbf{Experimental principle.}}
The studied quantum circuits realize a tunable single-channel quantum conductor in series with an adjustable resistance.
The suppression of the quantum conductor's conductance due to DCB is extracted either by increasing the temperature or voltage (exploiting the asymptotic vanishing of DCB), or by short-circuiting the series resistance in-situ using an on-chip field-effect switch. A quantum point contact (QPC) of adjustable width is used as a test-bed to emulate any short single-channel quantum conductor.

\vspace{\baselineskip}
{\noindent\textbf{Experimental implementation.}} The nano-circuits are tailored in a Ga(Al)As 2D electron gas and their conductance $G(V,T)=\partial I(V,T)/\partial V$ is measured at low temperatures, in a dilution refrigerator, using standard low-frequency lock-in techniques. The samples are constituted of three basic elements (see Fig.~1):

1) A single-channel quantum coherent conductor characterized by its transmission probability fully adjustable between $0$ and $1$. It is realized by a QPC formed by field effect in the 2D electron gas using a capacitively coupled metal split gate (at the bottom right in Fig.~1b, colorized in yellow) biased at a negative voltage.
The presence of well-defined plateaus at integer multiples of $1/R_\mathrm{q}$ in the QPC conductance $G$ versus split gate voltage ascertains that only one electronic quantum channel is partially open at time, with a transmission probability $\tau=R_\mathrm{q}G-n$ where $n$ corresponds to the number of fully open quantum channels. Note that in order to obtain a single partially open channel, spin degeneracy was broken with a large magnetic field perpendicular to the 2D electron gas that corresponds to the integer quantum Hall effect. Consequently, the electrical current propagates along several chiral edge channels, shown as lines in Fig.~1b with arrows indicating the propagation direction. In most cases, only the outer edge channel is partially reflected/transmitted at the studied QPC.

2) A dissipative environment characterized by a linear impedance $Z(\omega)$. It is realized by a resistance $R$, made either from a thin chromium wire deposited at the sample surface ($R=6.3$ and $80~\mathrm{k}\Omega$, see Fig.~1c) or from a second QPC (see Fig.~1b) set to the center of a resistance plateau $R_\mathrm{q}/n$ (in this case the QPC emulates a linear resistance unaffected by DCB \cite{flensberg1993,golubev2005dcb}). This resistance is in parallel with a small geometrical capacitance $C\approx2~\mathrm{fF}$. The studied QPC is connected to the series resistance through a small ohmic contact (OC in Fig.~1). The ohmic contact is necessary to establish a connection between the surface chromium wires and the 2D electron gas buried 85~nm below the surface. In the presence of a series QPC, it also plays the crucial role of an electron reservoir that separates the studied QPC and the series QPC into two distinct quantum conductors.

3) An on-chip switch to turn off DCB. The studied QPC can be isolated from the dissipative environment by diverting toward grounded electrodes the chiral edge channels that are returning from the small ohmic contact. This turns off the DCB suppression of the QPC's conductance. In practice, it is realized using additional metal gates close to the studied QPC (e.g. in Fig.~1b, the gate SW1 is used to short-circuit the dissipative environment of QPC1). In the schematic representation shown Fig.~1a, the switch is in parallel with the series resistance: when DCB is turned off, the studied QPC is directly voltage biased. Note that these extra gates allow us to characterize separately the different circuit elements, including the small ohmic contact.

\begin{figure}[!htbp]
\renewcommand{\figurename}{\textbf{Figure}}
\renewcommand{\thefigure}{\textbf{\arabic{figure}}}
\centering\includegraphics[width=1\columnwidth]{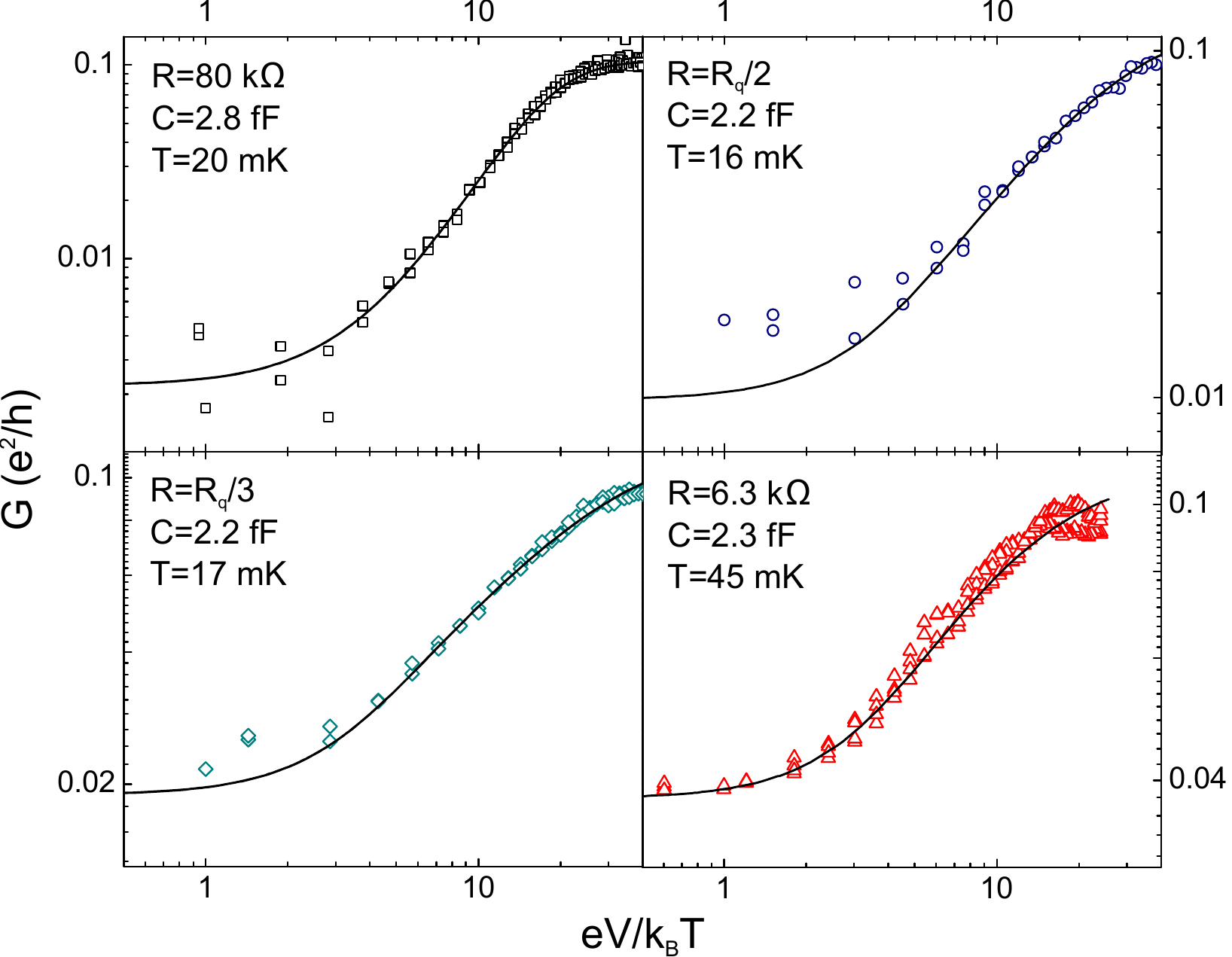}
\caption{\footnotesize
\textbf{Conductance suppression of a tunnel quantum channel} in a resistive environment. Symbols: low temperature conductance $G(V,T)=\partial I(V,T)/\partial V$ versus QPC DC voltage normalized by the temperature. Each panel corresponds to a QPC in the tunnel regime ($\Ginf\approx 0.1$) in series with a different on-chip resistance $R$. Continuous lines: prediction of the DCB theory for tunnel junctions using the separately determined values of $R$, $C$ and $T$ indicated in the panels.
\normalsize}
\label{fig-tunnel}
\end{figure}

\begin{figure}[!htb]
\renewcommand{\figurename}{\textbf{Figure}}
\renewcommand{\thefigure}{\textbf{\arabic{figure}}}
\centering\includegraphics [width=1\columnwidth]{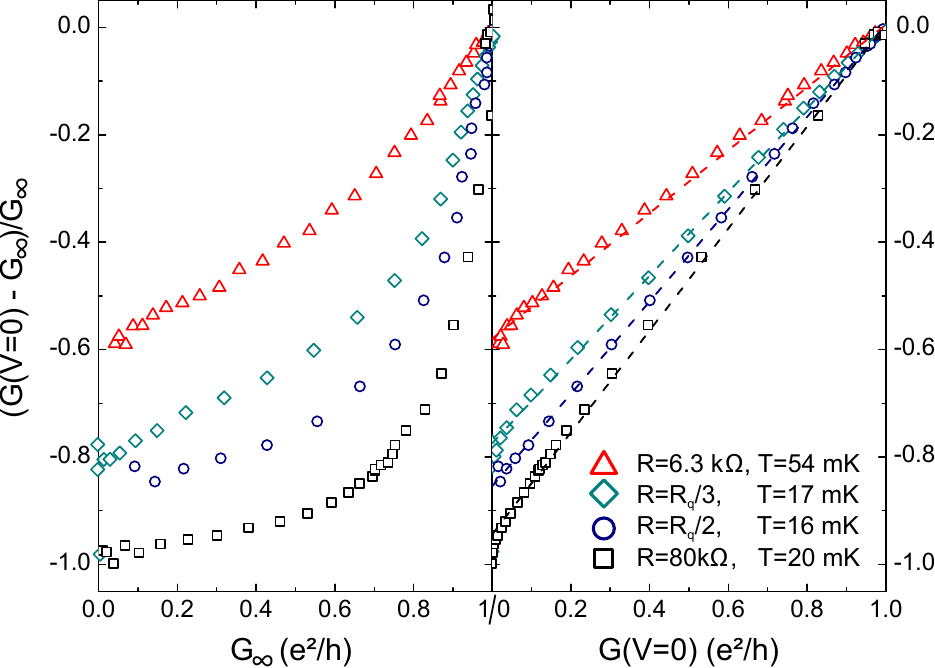}
\caption{\footnotesize
\textbf{Conductance suppression of an arbitrary quantum channel.} The relative suppression of the measured zero-bias QPC conductance $G(V=0,T)$ with respect to `intrinsic' quantum channel conductance $\Ginf$ is plotted as symbols versus $\Ginf$ (left panel) and $G(V=0,T)$ (right panel) for four series resistances $R=6.3~\mathrm{k}\Omega$, $R_\mathrm{q}/3$, $R_\mathrm{q}/2$ and $80~\mathrm{k}\Omega$. The straight dashed lines are guides for the eye.
\normalsize}
\label{fig-backaction}
\end{figure}

\vspace{\baselineskip}
{\noindent\textbf{Test of experimental procedure in the known tunnel junctions limit.}}
The experimental procedure to investigate DCB is tested by confronting the extracted conductance suppression of the QPC, set to a low transmission probability, with the known theoretical predictions for tunnel junctions (see e.g. Ref.~\onlinecite{ingold1992sct}).

Figure 2 shows as symbols the measured QPC conductance versus DC voltage $V$ for four different dissipative environments: two `macroscopic' on-chip chromium wires ($R=6.3~\mathrm{k}\Omega$ and $80~\mathrm{k}\Omega$) and two `mesoscopic' series QPCs set to a resistance plateau ($R=R_\mathrm{q}/2$ and $R=R_\mathrm{q}/3$). The conductance calculated using the DCB theory for tunnel junctions \cite{ingold1992sct} is shown as continuous black lines. The dissipative environment in the calculation is modeled by the schematic $R//C$ circuit shown Fig.~1a. Note that the only adjustable parameter is here the QPC's `intrinsic' conductance $\Ginf$ in absence of DCB, which is approximately given by the measured conductance at the highest applied voltages. The resistance $R$ injected in the calculation is measured directly on-chip, the capacitance $C$ corresponds to finite elements numerical simulations, and the temperature $T$ is set to that of the dilution fridge mixing chamber.

The good agreement in the tunnel regime between data and theory validates our experimental approach. It also shows that a QPC set to a well-defined resistance plateau $R=R_\mathrm{q}/n$ mimics a `macroscopic' linear resistance \cite{flensberg1993,golubev2005dcb}.

\vspace{\baselineskip}
{\noindent\textbf{Conductance suppression of a single-channel quantum conductor.}}
Figure 3 shows as symbols the relative suppression of the single-channel QPC conductance measured at zero DC voltage bias and low temperature for the same four environments tested in the tunnel regime. The quantum channel is characterized by its `intrinsic' conductance $\Ginf$, which is extracted by two methods: we either assign $\Ginf$ to the conductance measured with the dissipative environment short-circuited using the switch, or to the conductance measured at a large voltage bias where DCB corrections are small (see Supplementary Note~1 for further details).
The same data are plotted in the left panel of Fig.~3 versus $\Ginf$, and in the right panel versus the suppressed conductance $G(V=0,T)$.
The non-linear dependence exhibited in the left panel shows that the prediction derived in the weak DCB framework, of a relative conductance suppression proportional to $(1-R_\mathrm{q}\Ginf)$ for a single channel \cite{yeyati2001dcb,golubev2001dcb}, does not hold in the strong DCB regime.
Instead, we observe that the relative conductance suppression is proportional to $(1-R_\mathrm{q} G(V=0,T))$ at our experimental accuracy, as seen from the linear dependence exhibited in the right panel of Fig.~3. This result is a remarkable corroboration of the recent experimental finding in Ref.~\onlinecite{parmentier2011}, extending it to over more than one order of magnitude of the series resistance.

\begin{figure}[!htbp]
\renewcommand{\figurename}{\textbf{Figure}}
\renewcommand{\thefigure}{\textbf{\arabic{figure}}}
\centering\includegraphics[width=1\columnwidth]{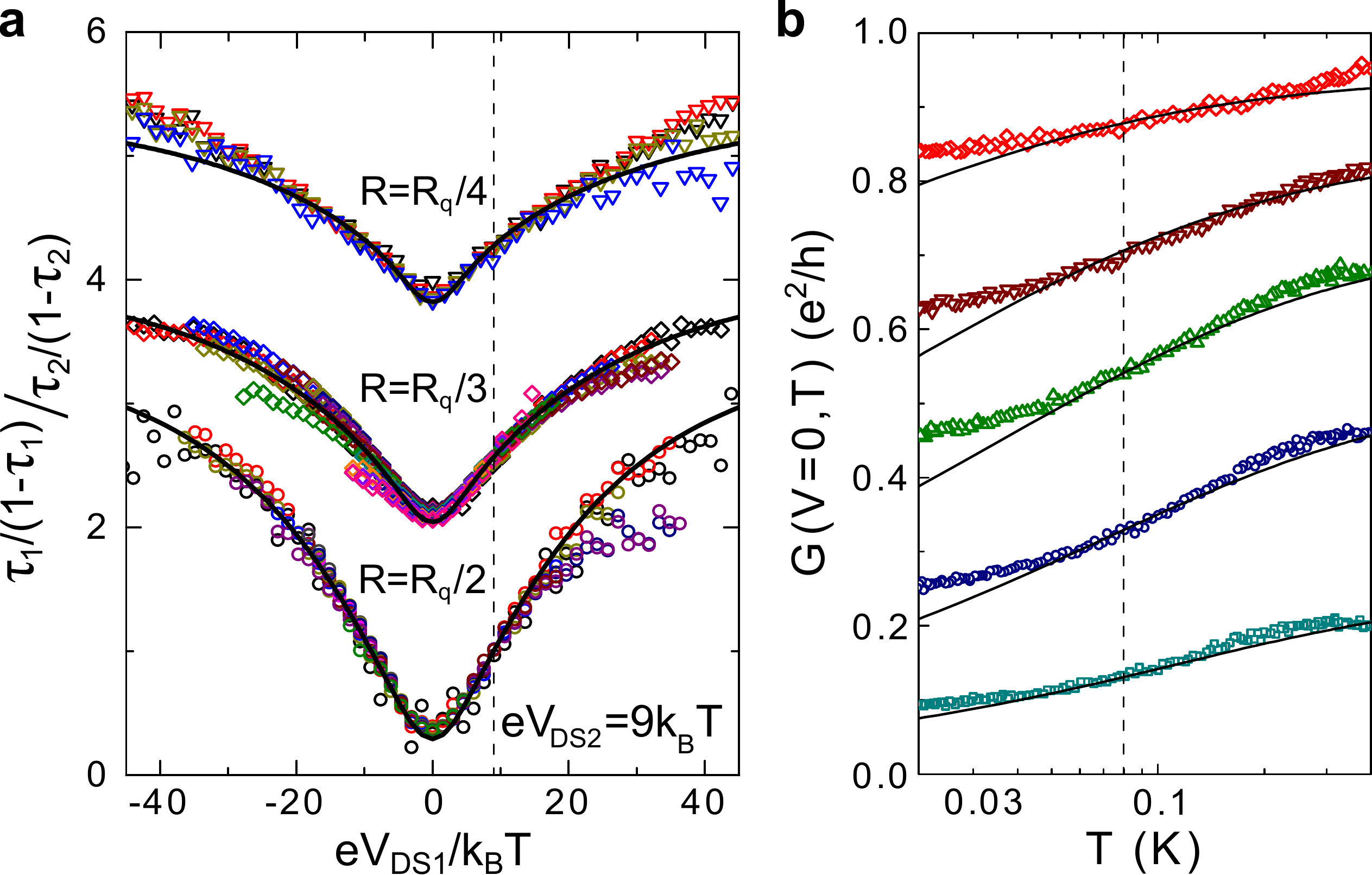}
\caption{\footnotesize
\textbf{Direct test of derived scaling law.} \textbf{(a)} Symbols: the measured low temperature QPC transmissions ($\tau_{1,2}=R_\mathrm{q}G(V_\mathrm{DS1,DS2},T)$) are recast as $\frac{\tau_1/(1-\tau_1)}{\tau_2/(1-\tau_2)}$ and plotted versus $V_\mathrm{DS1}$ for a fixed reference voltage $V_\mathrm{DS2}=9k_\mathrm{B}T$. For each of the three series resistances $R=R_\mathrm{q}/2$, $R_\mathrm{q}/3$ and $R_\mathrm{q}/4$ (corresponding datasets are shifted vertically for clarity), the QPC are tuned over a broad range of $\Ginf$, respectively $[0.06,0.95]e^2/h$, $[0.06,0.996]e^2/h$ and $[0.25,0.85]e^2/h$. Continuous lines: prediction of the phenomenological scaling law Eq.~\ref{EqScaling} using the corresponding $R$ and $C=2.0$~fF.
\textbf{(b)}, Symbols: zero-bias QPC conductance plotted versus temperature for the sample of series resistance $R=6.3~\mathrm{k}\Omega$, and with five different tuning of the QPC yielding different $\Ginf$. Continuous lines: Predictions of Eq.~\ref{EqScaling} calculated with $R=6.3~\mathrm{k}\Omega$, $C=2.3$~fF and using $T=80$~mK as the reference point.
\normalsize}
\label{fig-scaling}
\end{figure}

\vspace{\baselineskip}
{\noindent\textbf{Phenomenological expression and scaling law for the conductance.}}
The experimental observation of a relative conductance suppression approximately proportional to $(1-R_\mathrm{q} G)$ is highly non-trivial and has strong implications.

As shown in Ref.~\onlinecite{parmentier2011} (see also Supplementary Note~2), this finding implies a phenomenological expression for the conductance $G$ of a single channel embedded in a linear environment characterized by the impedance $Z(\omega)$:
\begin{eqnarray}
G(\Ginf,Z,V_\mathrm{DS},T) = \Ginf \frac{1+E_\mathrm{B}(Z,V_\mathrm{DS},T)}{1+R_\mathrm{q} \Ginf E_\mathrm{B}(Z,V_\mathrm{DS},T)},
\label{EqGeneral}
\end{eqnarray}
where $E_\mathrm{B}\equiv\lim_{\Ginf\rightarrow0}\frac{G-\Ginf}{\Ginf}$ is the relative conductance suppression in the tunnel regime, that can be calculated within the well-known DCB tunnel framework \cite{ingold1992sct}. Note that Eq.~\ref{EqGeneral} applies to short channels for which the energy $h/\tau_\mathrm{dwell}$, associated with the electronic dwell time $\tau_\mathrm{dwell}$, is larger than the other relevant energy scales (e.g. $eV_\mathrm{DS}$, $k_\mathrm{B}T$, $e^2/2C$). Indeed, even in absence of DCB, the conductance can change with the voltage and the temperature on the typical energy scale $h/\tau_\mathrm{dwell}$ (e.g. due to quantum interferences within the conductor). Moreover, a finite dwell time could result in a high-energy cutoff for the excited electromagnetic modes of the circuit, thereby reducing the overall conductance suppression due to DCB \cite{nazarov1991}.

The above phenomenological expression for $G$ requires the knowledge of the `intrinsic' conductance $\Ginf$, which is inconvenient when this quantity is not available. This is the case for the TLL predictions, due to the presence of a high-energy cutoff in the theory, or in experimental situations such as in Ref.~\onlinecite{mebrahtu2012} where the conductor's conductance changes significantly at high energy even in absence of DCB. It is therefore useful to remark that the above experimental finding can be recast as a scaling law relating the transmissions $\tau=R_\mathrm{q}G$ at two different energies without involving $\Ginf$ (see Supplementary Note~3):

\begin{eqnarray}
\frac{\tau_1/(1-\tau_1)}{\tau_2/(1-\tau_2)}=\frac{1+E_\mathrm{B}(Z,V_\mathrm{DS1},T_{1})}{1+E_\mathrm{B}(Z,V_\mathrm{DS2},T_{2})},
\label{EqScaling}
\end{eqnarray}
where $\tau_{1,2}\equiv G_{1,2}R_\mathrm{q}$ are the conductances, in units of conductance quantum, of the same single-channel quantum conductor in presence of DCB at the generator bias voltages $V_\mathrm{DS1}$ and $V_\mathrm{DS2}$, and at the temperatures $T_1$ and $T_2$.

A direct test of the scaling is displayed Fig.~4a, where the data obtained at $T=17$~mK for a wide range of $\Ginf$, from near tunnel to near full transmission, are recast following the above scaling law with a fixed reference voltage $V_\mathrm{DS2}=9k_\mathrm{B}T$. We observe that all the data corresponding to a given series resistance $R\in\{R_\mathrm{q}/2,R_\mathrm{q}/3,R_\mathrm{q}/4\}$ fall on top of each other, following the same black continuous line calculated with Eq.~\ref{EqScaling} without fit parameters. Note that we display only the data points on voltage ranges for which the separately measured energy dependency of the conductance in the absence of DCB is small (see Methods and Supplementary Note~4). Note also that for these series resistances, heating effects due to the voltage bias are negligible (see Supplementary Note~5).

In Figure 4b, we make use of the scaling law for the $R=6.3~\mathrm{k}\Omega$ series resistance's sample by taking as a reference point the QPC conductance at $T=80$~mK (dashed vertical line), which is high with respect to mismatches between electronic and mixing chamber temperatures, and low regarding temperature dependencies of the `intrinsic' transmission. We find that the measured conductances (symbols) plotted versus temperature obey the scaling law prediction of Eq.~\ref{EqScaling} (continuous lines) without any fit parameters and for a wide range of QPC tunings. Note that the discrepancies below 40~mK are possibly due to a higher electronic temperature for this set of data.

\begin{figure}[!htbp]
\renewcommand{\figurename}{\textbf{Figure}}
\renewcommand{\thefigure}{\textbf{\arabic{figure}}}
\centering\includegraphics[width=1\columnwidth]{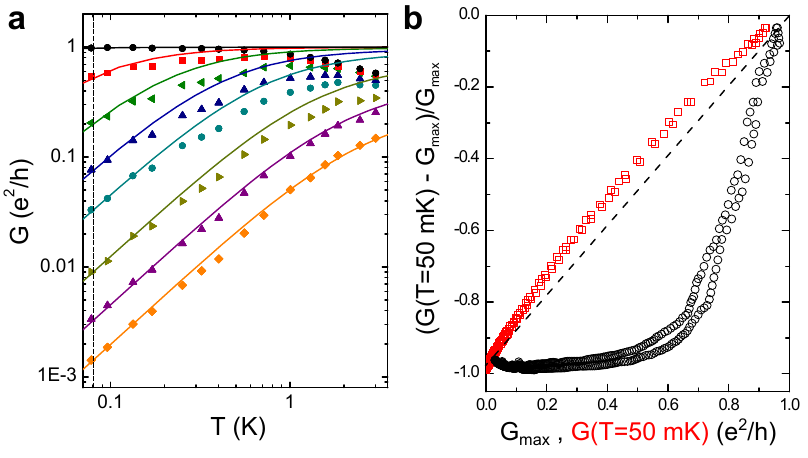}
\caption{\footnotesize
\textbf{Carbon nanotube resonant level in a resistive environment.}\textbf{(a)}, Symbols: zero-bias conductance of the carbon nanotube single-channel resonant level plotted  versus temperature (data extracted from Figure~4a in Ref.~\onlinecite{mebrahtu2012}), with each type of symbols corresponding to a different energy of the resonant level (tuned with a plunger gate, unity transmission corresponds to the Fermi energy \cite{mebrahtu2012}). Continuous lines: predictions calculated essentially without fit parameters with Eq.~\ref{EqScaling} using $R=0.75R_\mathrm{q}$, $C=0.07~$fF and a reference temperature $T=80$~mK indicated by a dashed line. \textbf{(b)}, The zero-bias relative conductance  suppression at $T=50$~mK with respect to the maximum conductance (extracted from the data shown in Fig.~2b of Ref.~\onlinecite{mebrahtu2012}) is shown as open symbols for different tunings of the resonant level position. The same data is plotted as circles versus the maximum conductance $G_\mathrm{max}$ ($\circ$) and as squares versus the conductance at $T=50$~mK ($\square$). Note that the appearance of two distinct curves, most visible when plotted versus $G_\mathrm{max}$, corresponds to the two sets of data obtained by shifting the resonant level in opposite direction from the Fermi energy. The straight dashed line is a guide for the eye.
\normalsize}
\label{fig-fink}
\end{figure}

\vspace{\baselineskip}
{\noindent\textbf{Comparison with a carbon nanotube resonant level.}} The conductance suppression for a single-channel quantum conductor inserted into a dissipative environment was recently measured on a markedly different physical system than in the present work, namely a resonant level within a carbon nanotube in series with on-chip resistances \cite{mebrahtu2012,mebrahtu2012b}. The fact that the resonant peaks are wide, much more than the temperature broadening \cite{mebrahtu2012}, implies that the energy associated with the dwell time across the nanotube is large with respect to temperature. According to the scattering approach \cite{landauer1975,anderson1980,buttiker1986}, such a resonant level realizes at low temperatures a short single-channel quantum conductor whose apparent complexity can be encapsulated in its `intrinsic' transmission probability. In Ref.~\onlinecite{mebrahtu2012}, the single-channel quantum conductor was tuned in-situ by adjusting the position of the resonant level with respect to the Fermi energy and by changing the symmetry of its coupling to the two connected leads.
As in the present work and in Ref.~\onlinecite{parmentier2011}, the authors of Refs~\onlinecite{mebrahtu2012,mebrahtu2012b} found that the conductance through the resonant level is strongly reduced at low temperatures only when it is tuned away from full transmission (the unitary limit) $G\simeq e^2/h$.

Here, we first compare the full temperature dependent conductance measured in Ref.~\onlinecite{mebrahtu2012} (symbols in Fig.~5a correspond to the data of Fig.~4a in Ref.~\onlinecite{mebrahtu2012}) with the prediction of the phenomenological scaling law Eq.~\ref{EqScaling} calculated essentially without fit parameters (continuous lines). The calculations are performed using the series resistance $R=0.75R_\mathrm{q}$ given in Ref.~\onlinecite{mebrahtu2012}, assuming a small parallel capacitance $C=0.07~$fF (there is little effect for realistic values $C\lesssim0.1~\mathrm{fF}$), and using $T_2=80~\mathrm{mK}$ as the reference temperature. We find a reasonable agreement between data and phenomenological scaling law Eq.~\ref{EqScaling} at low enough temperatures such that the symmetric resonant level aligned with the Fermi energy ($\bullet$) is perfectly transmitted. At higher temperatures, the conductance reduction in the symmetric case at the Fermi energy signals a transition toward either the sequential tunneling regime \cite{mebrahtu2012} or the long dwell-time regime with energy dependent transmissions \cite{mebrahtu2012b}, where the phenomenological expression Eq.~\ref{EqScaling} for short quantum conductors does not hold. Comparisons between the phenomenological scaling law Eq.~\ref{EqScaling} and the other data in Ref.~\onlinecite{mebrahtu2012}, as well as those in the new preprint Ref.~\onlinecite{mebrahtu2012b}, are available in Supplementary Note~6.

To expand further on the link with the present work, we plot in Fig.~5b the relative conductance reduction at $T=50~\mathrm{mK}$ with respect to the maximum conductance $\Gmax$ measured at the same gate voltage (extracted from Fig.~2b of Ref.~\onlinecite{mebrahtu2012}). The similarity with Fig.~3 is striking: whereas the relative conductance suppression plotted versus $\Gmax$ is markedly convex, it is close to a straight line when plotted as a function of $G(T=50~\mathrm{mK})$. Note that the discrepancy with a perfect linear behavior could be attributed to difference between $\Gmax$ and the `intrinsic' conductance $\Ginf$.

\vspace{\baselineskip}
{\noindent\textbf{Generalized mapping to a Tomonaga-Luttinger liquid.}}
The problem of a single-channel quantum conductor in a purely dissipative linear circuit, characterized by the series resistance $R$, can be mapped to that of a TLL of Luttinger interaction coefficient $1/(1+R/R_\mathrm{q})$ \cite{safi2004ohmic}. Remarkably, it can be shown that a frequency dependent circuit impedance corresponds to the more general problem of a 1D conductor with finite-range electron-electron interactions (see Supplementary Note~7). In the low energy limit, this more general model is known to reduce to a conventional TLL model with short-range interactions \cite{haldane1981}. Similarly, we establish here that realistic circuits with a high-frequency cut-off, e.g. capacitive as in Fig.~1a, can be mapped to a TLL.

More specifically, we consider the impact of the next orders in the Taylor series for the real part of a frequency dependent series impedance $\mathrm{Re}[Z(\omega)]=R+\sum_{n=1}^\infty R_n\left(\omega/\omega_\mathrm{Z}\right)^n$, where $\omega_\mathrm{Z}$ is the radius of convergence of the Taylor series expansion. As detailed in Supplementary Note~7, the electromagnetic environment shows up in the effective bosonic action, describing the electrical transport across the quantum conductor, as an additional quadratic term proportional to $\mathrm{Re}[Z(\omega)] |\hat Q(\omega)|^2$, where $\hat Q(t)$ is a bosonic field identifiable as the transferred charge. Then, by power counting arguments, we find that the leading term $R|\hat Q(\omega)|^2$ is most relevant at low-energy scales. Consequently this problem is described by the same action as for an impurity in a TLL. Note that the mapping applies provided the energy scales $k_\mathrm{B} T$ and $eV_\mathrm{DS}$ remain small compared to $\mathrm{min}[\hbar\omega_\mathrm{Z},\hbar\omega_\mathrm{F}]$. Here, $\hbar\omega_\mathrm{F}$ is a TLL cutoff that delimits both the validity of the short single-channel conductor approximation (limited by the finite dwell-time across the conductor and the energy barrier separating additional electronic channels) and of the linearization of the energy spectrum in the leads.

\vspace{\baselineskip}
{\noindent\textbf{Theoretical derivation of the phenomenological scaling law.}}
First, for purely dissipative circuits characterized by a small series resistance $Z(\omega)=R \ll R_\mathrm{q}$ but beyond the limit of weak conductance suppression, it was predicted both using a renormalization group approach \cite{kindermann2003fcs} and exploiting the mapping to TLL \cite{matveev1993tunnel1d,safi2004ohmic} that the energy dependent single-channel conductance $G=\tau / R_\mathrm{q}$ obeys the out-of-equilibrium flow equation ($k_\mathrm{B}T \ll eV_\mathrm{DS}$):
\begin{equation}\label{eq_r_1}
  \frac{d\,\tau (V_\mathrm{DS})}{d \log V_\mathrm{DS}}=\frac{2 R}{R_\mathrm{q}} \tau (V_\mathrm{DS})[1-\tau (V_\mathrm{DS})].
\end{equation}
This equation can be integrated between the applied generator voltages $V_\mathrm{DS1}$ and $V_\mathrm{DS2}$, which results in the same expression as the proposed phenomenological scaling law Eq.~\ref{EqScaling} for the corresponding limit of a purely dissipative circuit at $T=0$ (in which case \cite{panyukov1988,odintsov1988,nazarov1989,averin1989,devoret1990dcb,girvin1990dcb} $(1+E_\mathrm{B}) \equiv\lim_{\Ginf\rightarrow0}G/\Ginf \propto V_\mathrm{DS}^{2R/R_\mathrm{q}}$):
\begin{equation}\label{eq_r_int}
  \frac{\tau(V_\mathrm{DS1})/[1-\tau(V_\mathrm{DS1})]}{\tau(V_\mathrm{DS2})/[1-\tau(V_\mathrm{DS2})]}=\left(\frac{V_\mathrm{DS1}}{V_\mathrm{DS2}}\right)^{2R/R_\mathrm{q}}.
\end{equation}

Remarkably, we find here, using the thermodynamic Bethe ansatz solution of the impurity problem in a TLL at $k_\mathrm{B}T \ll eV_\mathrm{DS}$ \cite{fendley1995b}, that the same flow equation~\ref{eq_r_1} and, consequently, the phenomenological scaling law Eq.~\ref{EqScaling} are obeyed beyond the limit $R \ll R_\mathrm{q}$: as detailed in Supplementary Note~8, we obtain the flow equation~\ref{eq_r_1} exactly at $R=R_\mathrm{q}$ within the full generalized validity domain of the mapping ($e V_\mathrm{DS}$ below $\mathrm{min}[\hbar\omega_\mathrm{Z},\hbar\omega_\mathrm{F}]$). Note that for $R=R_\mathrm{q}$, corresponding to a Luttinger interaction coefficient $K=1/2$, the same conclusions can be reached by an alternative theoretical approach referred to as the refermionization procedure \cite{guinea1985,matveev1995,chamon1996}. We also obtain Eq.~\ref{eq_r_1} for arbitrary values of $R$ in the low $V_\mathrm{DS}$ limit corresponding to small values of the suppressed transmission $\tau (V_\mathrm{DS}) \ll 1$ (even if the corresponding $\tau_\infty$ is close to unity).

\begin{figure}
\renewcommand{\figurename}{\textbf{Figure}}
\renewcommand{\thefigure}{\textbf{\arabic{figure}}}
\includegraphics[width=1\columnwidth]{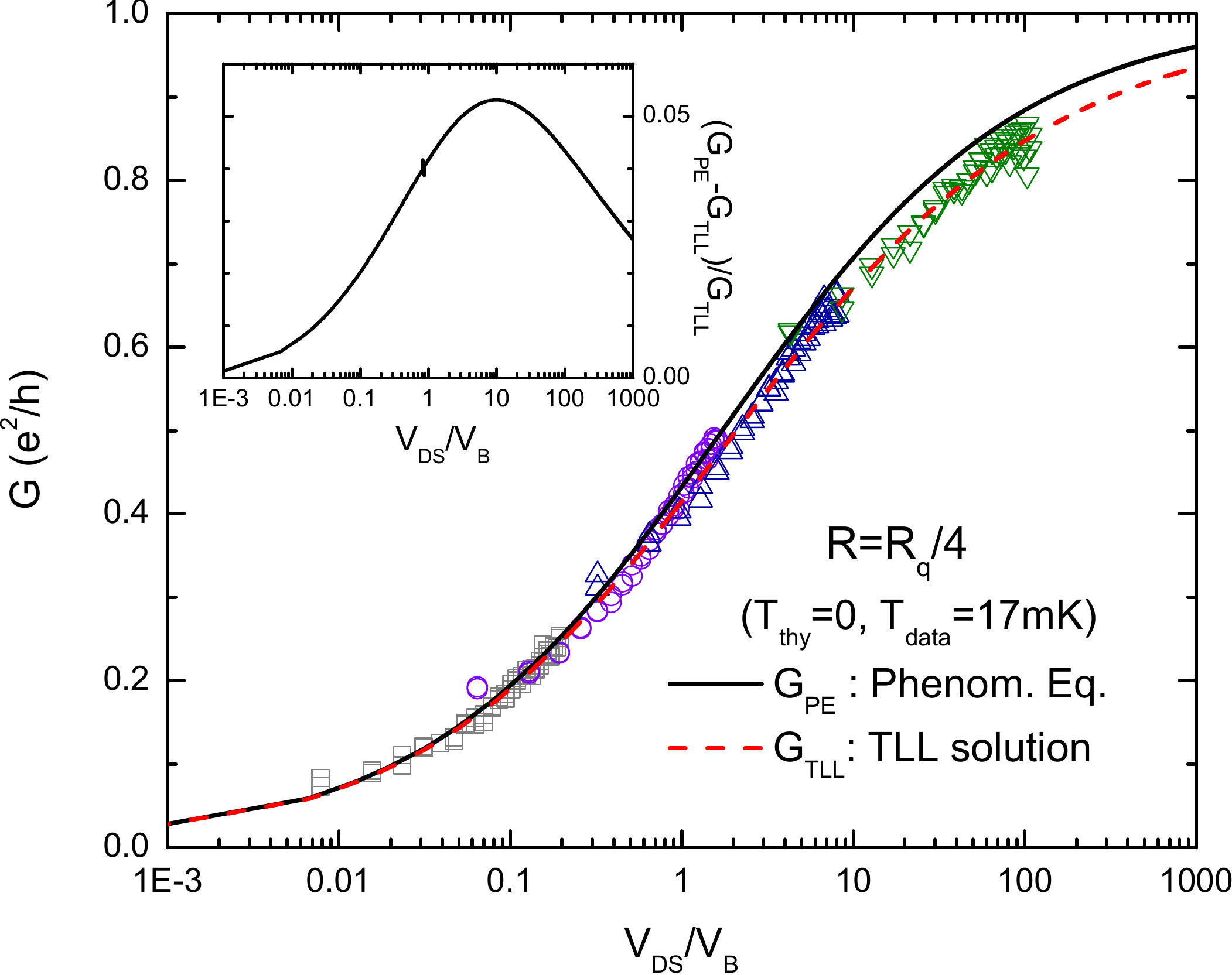}
\caption{\textbf{Comparison TLL universal conductance curve-phenomenological scaling law-data} for $R=R_\mathrm{q}/4$, beyond the theoretically established validity of the phenomenological scaling law. Red dashed line: universal conductance scaling curve $G_\mathrm{TLL}(V_\mathrm{DS}/V_\mathrm{B})$ computed from the TLL thermodynamic Bethe ansatz solution at $T=0$ \cite{fendley1995,fendley1995b} (see also Supplementary Note~8). Black straight line: conductance $G_\mathrm{PE}$ predicted by the phenomenological scaling law given by Eq.~\ref{eq_r_int} for a pure series resistance at $T=0$. The full conductance curve $G_\mathrm{PE}(V_\mathrm{DS}/V_\mathrm{B})$ was calculated using the single reference point $G_\mathrm{PE}=G_\mathrm{TLL}$ at $V_\mathrm{DSref}/V_\mathrm{B}=0.0005$. Symbols: four data sets measured at $T=17$~mK in the presence of a series resistance $R=R_\mathrm{q}/4$. The corresponding scaling voltages $V_\mathrm{B}=330~\mu$V ($\square$), $40~\mu$V ($\circ$), $8~\mu$V ($\triangle$) and $0.6~\mu$V ($\triangledown$) are obtained by matching $G_\mathrm{TLL}(V_\mathrm{DSref}/V_\mathrm{B})$ with the conductance measured at the reference point $V_\mathrm{DSref}=9k_\mathrm{B}T\simeq 13~\mu$V (note that a better agreement data-phenomenological scaling law would have been obtained using instead $G_\mathrm{PE}(V_\mathrm{DSref}/V_\mathrm{B})$). Inset: relative deviations of the conductance calculated using the phenomenological expression Eq.~\ref{EqScaling} with respect to the universal conductance scaling curve $(G_\mathrm{PE}-G_\mathrm{TLL})/G_\mathrm{TLL}$.\label{fig:scali}}
\end{figure}

\vspace{\baselineskip}
{\noindent\textbf{Direct conductance comparison TLL-phenomenological scaling law.}}
Now that the validity of the proposed phenomenological scaling law Eq.~\ref{EqScaling} is theoretically established in the three different limits $\tau(V_\mathrm{DS}) \ll 1$, $R \ll R_\mathrm{q}$ and $R=R_\mathrm{q}$, we confront its predictions at intermediate values of $R$ with numerical evaluations of the exact TLL universal conductance curve.

In figure~6, we display such a comparison for the intermediate series resistance $R=R_\mathrm{q}/4$ on the full range of single-channel conductances.
The out-of-equilibrium ($k_\mathrm{B} T \ll eV_\mathrm{DS}$) TLL prediction for the conductance in the presence of an impurity follows a universal scaling curve $G_\mathrm{TLL}(V_\mathrm{DS}/V_\mathrm{B})$, with $V_\mathrm{B}$ a scaling voltage encapsulating the impurity potential. As detailed in Supplementary Note~8, this conductance can be numerically computed using the exact thermodynamic Bethe ansatz solution \cite{fendley1995,fendley1995b} (red dashed line in Fig.~6). Although the conductance curve depends on the TLL interaction coefficient $K=1/(1+R/R_\mathrm{q})$, it is universal in the sense that the same curve applies for an arbitrary local impurity. Note that there is no universal relation between $V_\mathrm{B}$ and the `intrinsic' transmission probability $\tau_\infty$ of the corresponding DCB problem. For instance, such a relation would depend on the specific high-frequency behavior of $Z(\omega)$.
The conductance predicted by the phenomenological scaling law Eq.~\ref{EqScaling} is shown as a continuous black line in Fig.~6. In the corresponding limit of a pure series resistance $Z(\omega)=R_\mathrm{q}/4$ at $k_\mathrm{B} T \ll eV_\mathrm{DS}$, it takes the simple analytical form given by Eq.~\ref{eq_r_int}. The single reference point used in the phenomenological scaling law is the TLL conductance prediction at a very low voltage bias $V_\mathrm{DSref}/V_\mathrm{B}=0.0005$, where it is theoretically established that both predictions match. Note that there is no need to fix the scaling voltage $V_\mathrm{B}$ here since only the voltage ratio with respect to $V_\mathrm{DSref}$ is needed in Eq.~\ref{eq_r_int}.

We find a good quantitative agreement between TLL and phenomenological scaling law predictions on the full range of bias voltages and conductances, with relatively small deviations appearing at large bias voltages ($\lesssim 5~\%$, see inset of Fig.~6). This good agreement corroborates the predicted mapping DCB-TLL. It is noteworthy that the phenomenological scaling law Eq.~\ref{EqScaling} encompasses arbitrary linear circuit impedances $Z(\omega)$, beyond the limit of a series resistance $Z(\omega)\simeq R$, suggesting a possible generalization of the TLL predictions.

\vspace{\baselineskip}
{\noindent\textbf{Experimental test of the predicted mapping TLL-DCB at $R=R_\mathrm{q}/4$.}}
The most straightforward experimental test consists in the direct comparison of the conductance data with the universal conductance scaling curve $G_\mathrm{TLL}(V_\mathrm{DS}/V_\mathrm{B})$.
Figure 6 displays such a comparison for the series resistance $R=R_\mathrm{q}/4$.
Four data sets of the conductance measured at $T=17~$mK, each corresponding to a different tuning of the QPC embedded in the same $R=R_\mathrm{q}/4$ environment, are shown as symbols. For each data set, the value of $V_\mathrm{B}$ is fixed by matching the measured conductance at a single arbitrary reference voltage $V_\mathrm{DSref}=9k_\mathrm{B}T\simeq 13~\mu$V with the TLL prediction for the conductance $G_\mathrm{TLL}(V_\mathrm{DSref}/V_\mathrm{B})$. This gives $V_\mathrm{B}=330~\mu$V ($\square$), $40~\mu$V ($\circ$), $8~\mu$V ($\triangle$) and $0.6~\mu$V ($\triangledown$). Note first that the lowest voltage in each data set corresponds to approximately $3k_\mathrm{B}T/e$, thereby minimizing the effect of the finite experimental temperature. Note also that the highest voltage in each data set $V_\mathrm{DS}\approx 65~\mu$V is smaller than $h/eRC \approx 300~\mu$V, which limits the contribution of the experimental short-circuit capacitance $C\simeq 2~$pF to the series impedance $Z(\omega)\simeq R_\mathrm{q}/4$.

We observe that the conductance data closely obey the TLL predictions over the full range of single-channel conductances and over four order of magnitudes of $V_\mathrm{DS}/V_\mathrm{B}$. This observation constitutes a direct experimental demonstration that the transport across a single-channel quantum conductor embedded in a dissipative environment can be mapped onto collective Tomonaga-Luttinger liquid behaviors.

\vspace{\baselineskip}
{\large\noindent\textbf{Discussion}}\\
The present work is at the crossroad of two seemingly distinct phenomena namely, on the one hand, the Tomonaga-Luttinger physics of interacting 1D conductors and, on the other hand, the different set of quantum laws of electricity when distinct quantum coherent conductors are assembled into a circuit. By advancing and confronting both the experimental and theoretical aspects, we have established the predicted link between these two phenomena for the basic class of mesoscopic circuits constituted by a short single-channel quantum conductor in series with a linear resistance. This opens the path to using electronic circuits as test-beds for Luttinger physics, and also advances our understanding of the quantum laws of electricity through the powerful TLL theoretical framework. In particular, important insight may be obtained in the investigation of the direct link between suppressed conductance and quantum shot noise, that is expected to hold even in the regime of strong conductance suppression \cite{kindermann2003fcs,safi2004ohmic} (see also Supplementary Note~9). An important outcome of the present work is that we strongly consolidate, delimit the validity, and grasp the significance of the generalized phenomenological expression Eq.~\ref{EqGeneral} for the conductance of an arbitrary short quantum channel in a linear environment. From an experimental standpoint, its validity is demonstrated for a wide range of circuit impedances, and is found in good agreement with the data of Refs~\onlinecite{mebrahtu2012,mebrahtu2012b} obtained on a different physical system, a carbon nanotube resonant level. From a theoretical standpoint, the equivalent scaling law  Eq.~\ref{EqScaling} is derived for the suppressed conductance in various limits, in particular for a series resistance $R=R_\mathrm{q}$. We also find that relatively small deviations exist in intermediate regimes. These results are not only of fundamental importance; the knowledge of the different quantum laws of electricity with coherent conductors has also direct implications for the quantum engineering of future nanoelectronic devices.

\vspace{\baselineskip}
{\large\noindent\textbf{Methods}}\\
{\noindent\textbf{Measured samples.}}
The samples are nanostructured by standard e-beam lithography in a 94~nm deep GaAs/Ga(Al)As two-dimensional electron gas of density $2.5\times10^{15}~\mathrm{m}^{-2}$ and mobility $55~\mathrm{m}^2\mathrm{V}^{-1}\mathrm{s}^{-1}$.

\vspace{\baselineskip}
{\noindent\textbf{Experimental setup.}}
The measurements were performed in a dilution refrigerator with a base temperature of $T=16$~mK. All measurement lines were filtered by commercial $\pi$-filters at the top of the cryostat. At low temperature, the lines were carefully filtered and thermalized by arranging them as 1~m long resistive twisted pairs ($300~\Omega /$m) inserted inside 260~$\mu$m inner diameter CuNi tubes, which were tightly wrapped around a copper plate screwed to the mixing chamber. The samples were further protected from spurious high energy photons by two shields, both at the mixing chamber temperature.

\vspace{\baselineskip}
{\noindent\textbf{Measurement techniques.}}
The differential conductance measurements were performed using standard lock-in techniques at frequencies below 100~Hz. To avoid sample heating, the AC excitation voltages across the sample were smaller than $k_\mathrm{B}T/e$. The sample was current biased by a voltage source in series with a $10~$M$\Omega$ or $100~$M$\Omega$ polarization resistance. The bias current applied to the drain was converted on-chip into a fixed $V_\mathrm{DS}$, independent of the QPC conductance, by taking advantage of the well defined quantum Hall resistance to ground of the drain electrode ($R_\mathrm{q}/n$ at filling factor $\nu = n$). Similarly, the current across each component (QPCs, switches) is obtained by converting the voltage measured with the amplifiers represented as triangles in Figure~1 using the $R_\mathrm{q}/n$ quantum Hall resistance. The conductances of the QPC, switch and series chromium wires or series QPC were obtained separately by three point measurements. For all the samples, we used cold grounds directly connected to the mixing chamber of the dilution refrigerator.

\vspace{\baselineskip}
{\noindent\textbf{Test of the small ohmic contacts.}}
The electrical connection between the small ohmic contact (labeled OC in article Fig.~1b) and the buried 2D electron gas was tested with both the QPCs and the switches set in the middle of the very large and robust conductance plateau $G=2/R_\mathrm{q}$. Assuming that the two outer edge channels are fully transmitted across QPCs and switches, and that the inner channels are fully reflected, we find for all samples that the reflection of each of the two outer edge channels on the small ohmic contact is smaller than $0.01$.

\vspace{\baselineskip}
{\noindent\textbf{Energy dependences of the `intrinsic' conductance.}}
A coherent conductor may present energy dependences in its `intrinsic' conductance $\Ginf$ associated with e.g. a finite dwell time. In the case of QPCs these often result from nearby defects. These energy dependences add up with the DCB energy dependence, which makes the extraction of the DCB signal as a function of voltage and temperature more difficult. In Supplementary Note~4, we illustrate the energy behavior of the QPCs with the electromagnetic environment short-circuited and explain how we deal with the energy dependences of $\Ginf$ in the present work.


\vspace{\baselineskip}
{\noindent\textbf{Acknowledgments}}\\
The authors gratefully acknowledge Y.~Nazarov, the Quantronics group and E.~Sukhorukov for discussions, F.~Lafont for his contribution to the experiment, and L.~Couraud, D.~Mailly, H.~le~Sueur and C.~Ulysse for their inputs in the nano-fabrication. This work was supported by the ERC (ERC-2010-StG-20091028, \#259033) and the ANR (ANR-09-BLAN-0199).

\vspace{\baselineskip}
{\noindent\textbf{Author contributions}}\\
Experimental work and analysis: S.J., F.D.P., A.A. and F.P.; heterojunction growth: A.C. and U.G.; nanofabrication: F.D.P and A.A; TLL theory: M.A., F.P. and I.S.; manuscript preparation: A.A., S.J., F.D.P., U.G., M.A., I.S. and F.P.; project planning and supervision: A.A. and F.P.

\vspace{\baselineskip}
{\noindent\textbf{Competing financial interests:} The authors declare no competing financial interests.}

\end{document}



\title{Supplementary Information for \\`Tomonaga-Luttinger physics in electronic quantum circuits'}

\author{S. Jezouin}
\affiliation{CNRS / Univ Paris Diderot (Sorbonne Paris Cit\'e), Laboratoire de Photonique et de Nanostructures
(LPN), route de Nozay, 91460 Marcoussis, France}
\author{M. Albert}
\affiliation{CNRS / Univ Paris Sud, Laboratoire de Physique des Solides
(LPS), 91405 Orsay, France}
\author{F.D. Parmentier}
\affiliation{CNRS / Univ Paris Diderot (Sorbonne Paris Cit\'e), Laboratoire de Photonique et de Nanostructures
(LPN), route de Nozay, 91460 Marcoussis, France}
\author{A. Anthore}
\affiliation{CNRS / Univ Paris Diderot (Sorbonne Paris Cit\'e), Laboratoire de Photonique et de Nanostructures
(LPN), route de Nozay, 91460 Marcoussis, France}
\author{U. Gennser}
\affiliation{CNRS / Univ Paris Diderot (Sorbonne Paris Cit\'e), Laboratoire de Photonique et de Nanostructures
(LPN), route de Nozay, 91460 Marcoussis, France}
\author{A. Cavanna}
\affiliation{CNRS / Univ Paris Diderot (Sorbonne Paris Cit\'e), Laboratoire de Photonique et de Nanostructures
(LPN), route de Nozay, 91460 Marcoussis, France}
\author{I. Safi}
\affiliation{CNRS / Univ Paris Sud, Laboratoire de Physique des Solides
(LPS), 91405 Orsay, France}
\author{F. Pierre}
\affiliation{CNRS / Univ Paris Diderot (Sorbonne Paris Cit\'e), Laboratoire de Photonique et de Nanostructures
(LPN), route de Nozay, 91460 Marcoussis, France}

\maketitle

\section{Supplementary Figures}
\begin{figure}[h]
\renewcommand{\figurename}{\textbf{Supplementary Figure}}
\renewcommand{\thefigure}{\textbf{S\arabic{figure}}}
\centering\includegraphics[width=0.75\columnwidth]{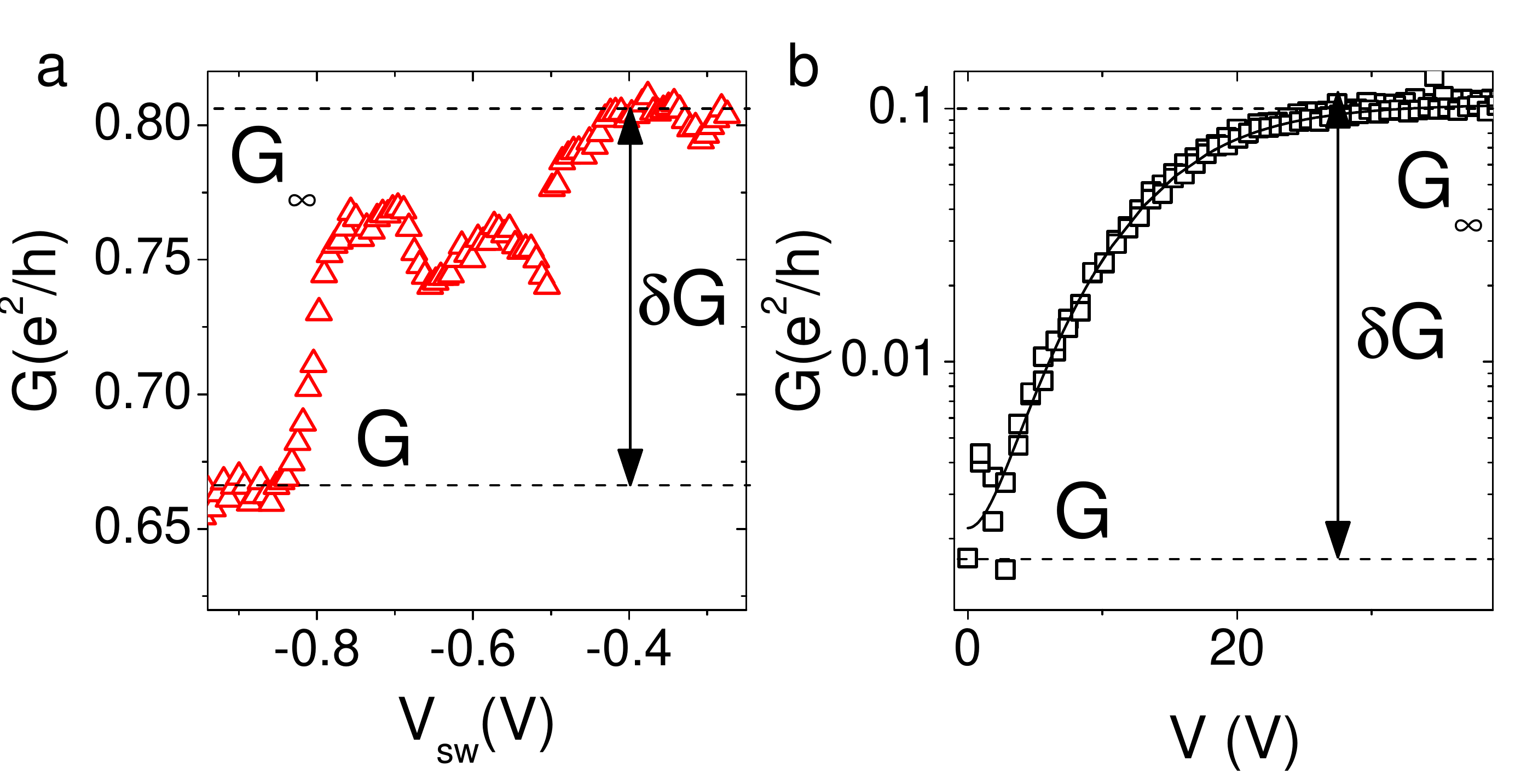}
\caption{\textbf{Extraction of the `intrinsic' conductance $\Ginf$}. (a) QPC conductance versus the switch voltage $\VSW$ for the $R=6~\mathrm{k}\Omega $ sample. (b) QPC conductance versus the DC voltage $V$ for the $R=80~\mathrm{k}\Omega $ sample.
}
\label{fig-SIfig1}
\end{figure}
\begin{figure}
\renewcommand{\figurename}{\textbf{Supplementary Figure}}
\renewcommand{\thefigure}{\textbf{S\arabic{figure}}}
\centering\includegraphics[width=0.8\columnwidth]{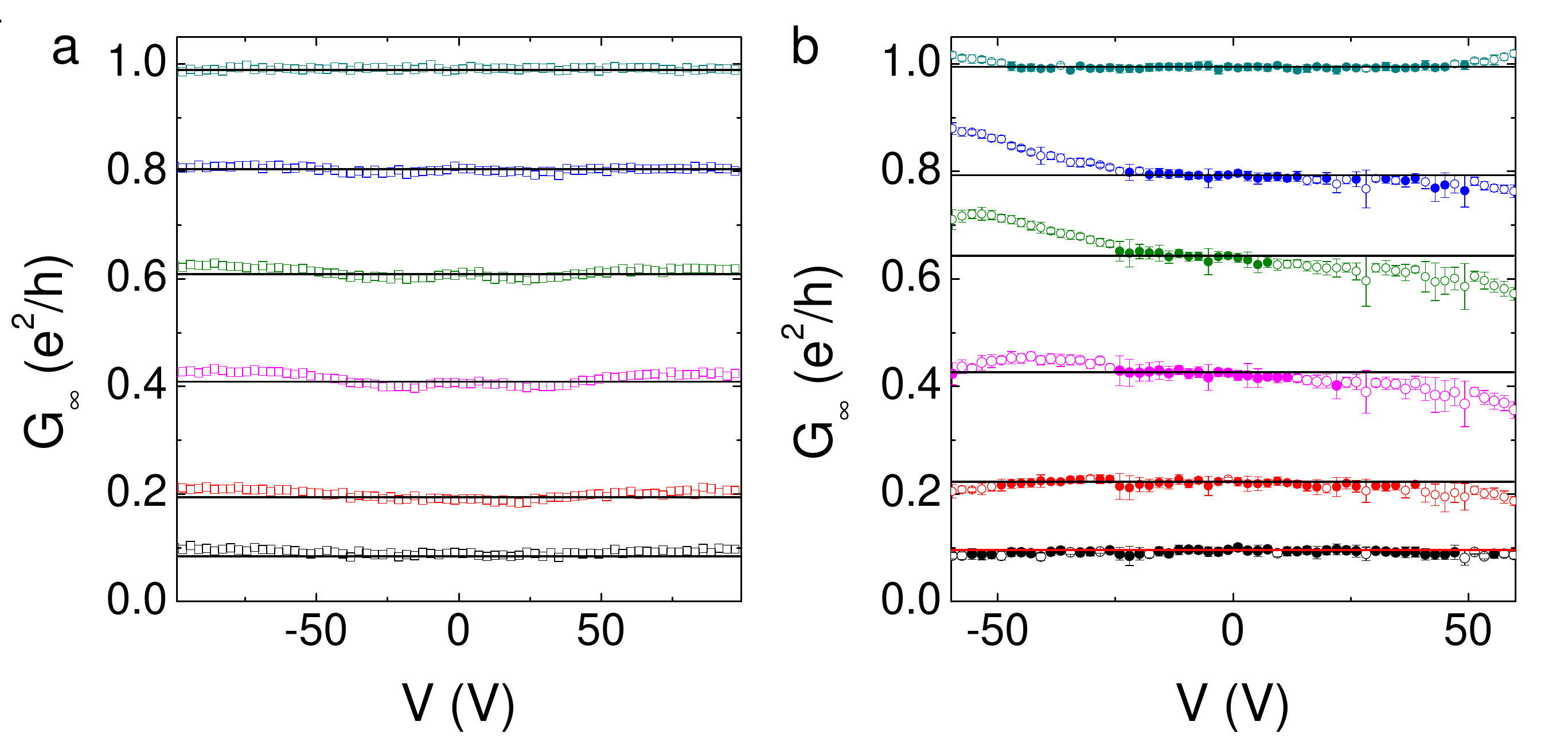}
\caption{\textbf{Energy dependences of the `intrinsic' conductance}. (a) QPC conductance versus QPC DC voltage $V$ for the $R=80~\mathrm{k}\Omega$ sample. The horizontal lines are the `intrinsic' conductance measured at $V=0$. (b) Symbols: QPC conductance versus QPC DC voltage $V$ for the $R_\mathrm{q}/2$ sample. Full symbols: points that were kept for plotting the scaling law of article figure~4a. Error bars are statistical errors deduced from $18$ measurements. The horizontal lines are the `intrinsic' conductances measured at $V=0$.
}
\label{fig-SIfig2}
\end{figure}

\begin{figure}
\renewcommand{\figurename}{\textbf{Supplementary Figure}}
\renewcommand{\thefigure}{\textbf{S\arabic{figure}}}
\centering\includegraphics[width=0.5\columnwidth]{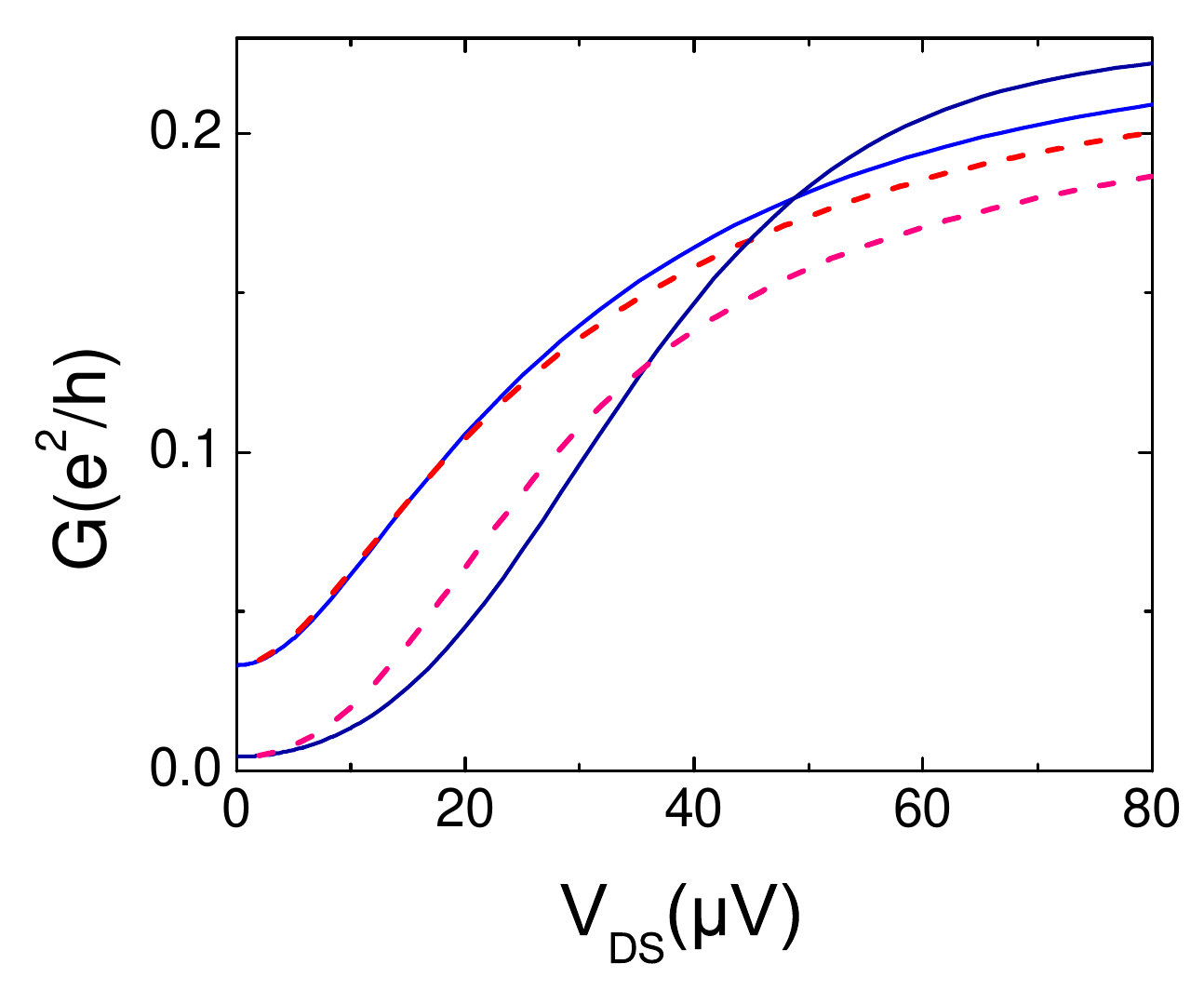}
\caption{\footnotesize
\textbf{Effect of heating at finite bias.} Continuous lines: calculated conductance $G=\frac{dI}{dV}$ without taking into account heating effect, using $\tauinf=0.23$, $C=2~$fF, $T=25~$mK and either $R=80~\mathrm{k}\Omega$ (bottom line) or $R=R_\mathrm{q}/2$ (top line). Dashed lines~: calculated conductance including heating effects with the simple model in the supplementary information of [43].
\normalsize}
\label{fig-SIfig3}
\end{figure}
\begin{figure}
\renewcommand{\figurename}{\textbf{Supplementary Figure}}
\renewcommand{\thefigure}{\textbf{S\arabic{figure}}}
\centering\includegraphics[width=0.8\columnwidth]{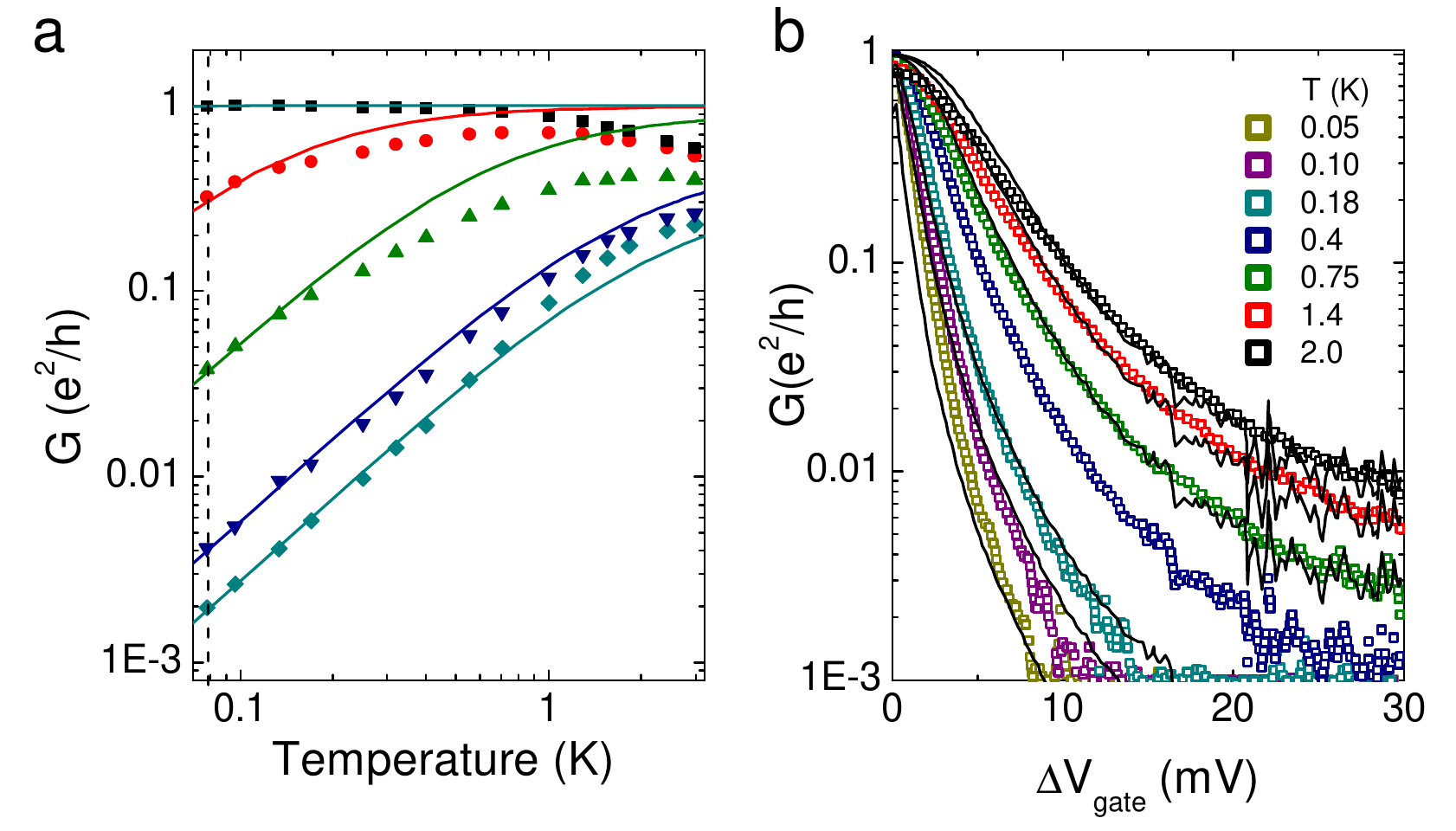}
\caption{\footnotesize
\textbf{Carbon nanotube resonant level in a resistive environment.} \textbf{(a)}  Symbols: Zero bias conductance of the carbon nanotube single-channel resonant level plotted  versus temperature (data extracted from Figure~3a in [44]), with each type of symbols corresponding to different degrees of asymmetry. Continuous lines: predictions calculated without fit parameters with supplementary Eq.~\ref{EqSI} using $R=0.75R_\mathrm{q}$, $C=0.07~\mathrm{fF}$ and as reference points the conductances measured at $T_\mathrm{ref}=80$~mK. \textbf{(b)} Symbols: Zero bias conductance of the carbon nanotube single-channel resonant level, tuned to a symmetric coupling with the dissipative leads, plotted versus $\Delta V_{\mathrm{gate}}$, the variation in the plunger gate voltage controlling the energy of the resonant level with respect to the center of the peak (data extracted from Figure~2b in [44]). Each type of symbols corresponds to a different temperature. Continuous lines: predictions calculated without fit parameters with supplementary Eq.~\ref{EqSI} using $R=0.75R_\mathrm{q}$, $C=0.07~\mathrm{fF}$, and as reference points the measured $G(\Delta V_{\mathrm{gate}},T_\mathrm{ref}=0.4~\mathrm{K})$.
\normalsize}
\label{fig-SIfig4}
\end{figure}

\begin{figure}
\renewcommand{\figurename}{\textbf{Supplementary Figure}}
\renewcommand{\thefigure}{\textbf{S\arabic{figure}}}
\centering\includegraphics[width=0.8\columnwidth]{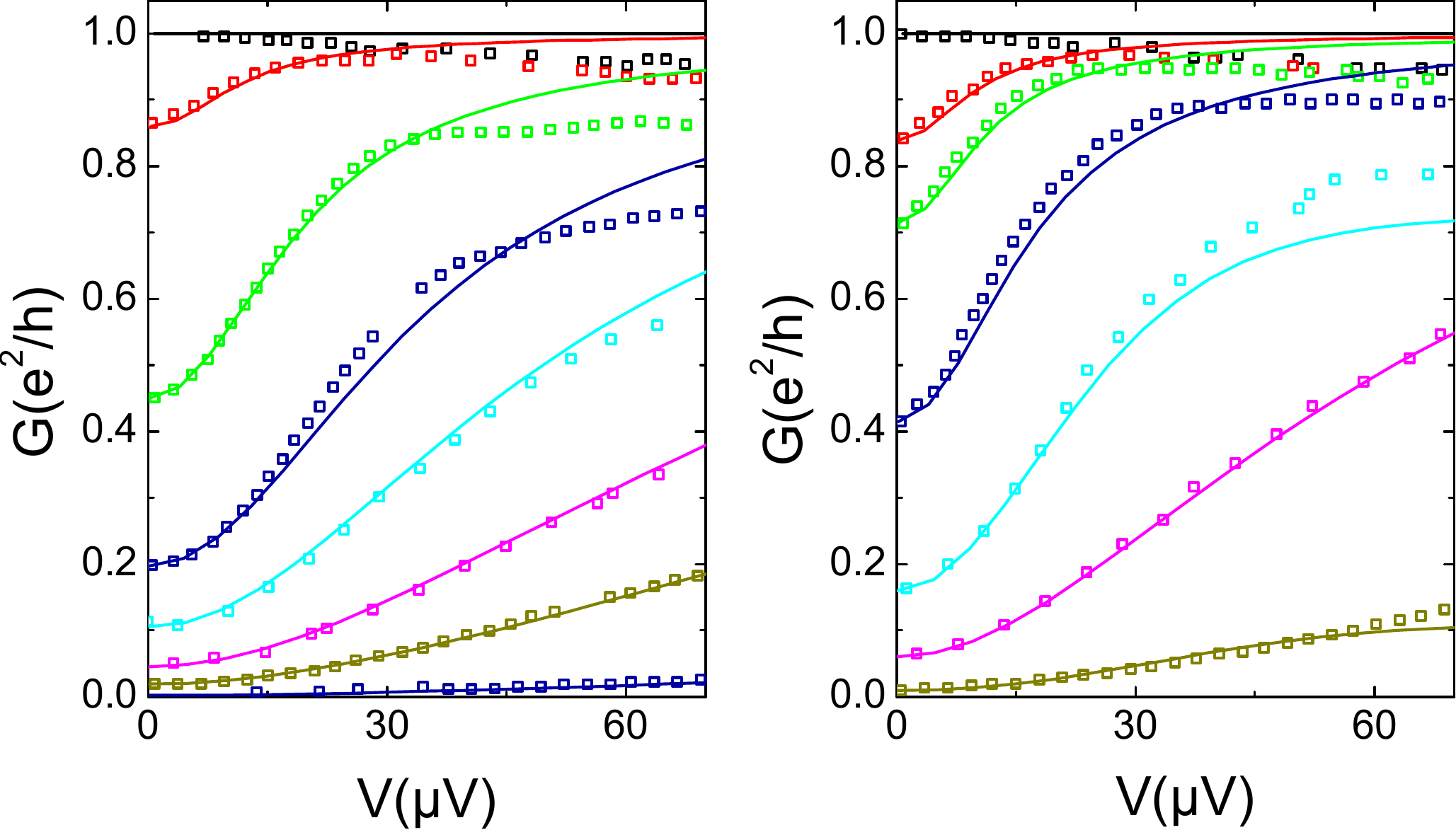}
\caption{\footnotesize
\textbf{Carbon nanotube resonant level in a resistive environment.} Symbols: Conductance of the carbon nanotube single-channel resonant level plotted  versus voltage (data extracted from Figure~S1 in the supplementary information of [45]), with each type of symbols corresponding to a different tuning of the single-channel resonant level. In the left panel, the single-channel was tuned by changing the asymmetric coupling to the dissipative leads, with the resonant level remaining at the Fermi energy. In the right panel, the single-channel was tuned by changing the position of the resonant level coupled symmetrically to the dissipative leads. Continuous lines: predictions calculated essentially without fit parameters with supplementary Eq.~\ref{EqSI}, using $R=0.75R_\mathrm{q}$, $C=0.07~\mathrm{fF}$, $T=50 (45)~\mathrm{mK}$ for the left (right) panel, and using as a reference point in supplementary Eq.~\ref{EqSI} the conductance measured at zero bias voltage.
\normalsize}
\label{fig-SIfig5}
\end{figure}
\begin{figure}[h]
\renewcommand{\figurename}{\textbf{Supplementary Figure}}
\renewcommand{\thefigure}{\textbf{S\arabic{figure}}}
\centering\includegraphics[width=0.5\columnwidth]{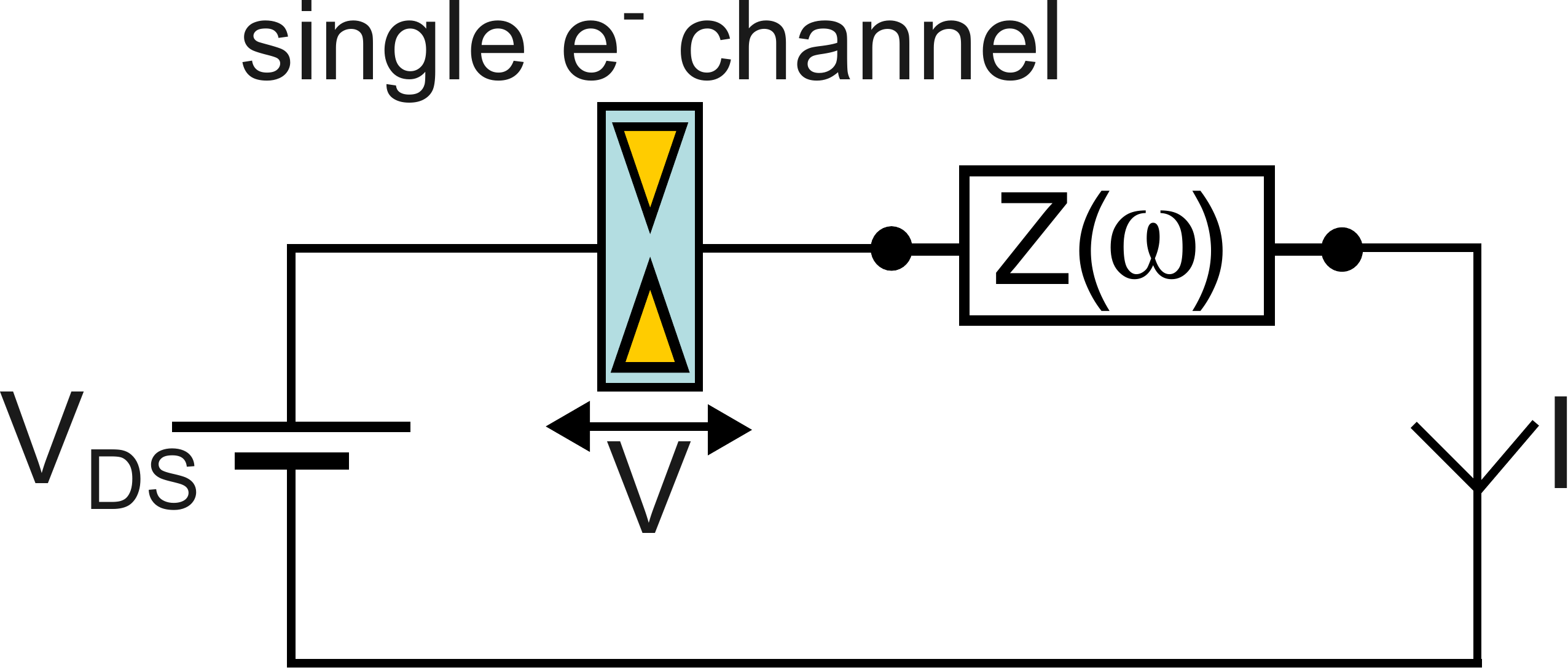}
\caption{\footnotesize
\textbf{Schematic circuit considered in the mapping to TLL.}
\normalsize}
\label{fig-SIfig6}
\end{figure}

\clearpage

\section{Supplementary Notes}

{\large\noindent\textbf{Supplementary Note 1: Extraction of the `intrinsic' conductance of the quantum coherent conductor}}\\

To extract the conductance reduction of the quantum coherent conductor due to dynamical Coulomb blockade (DCB), shown on the article Figure~3, one needs to extract its `intrinsic' conductance (i.e. the conductance $\Ginf$ in absence of DCB.) Two methods were used for the QPC : we either assign $\Ginf$ to the conductance measured with the dissipative environment short-circuited using the switch, or to the conductance measured at a large bias voltage where DCB corrections are small (typically in the range $50-100~\mu$V). Both methods have advantages and drawbacks detailed in this note.

Supplementary figure~S1 illustrates the two different methods. In the `switch method' (left panel), the gate voltage $\VSW$ controlling the switch is modified in order to close it. For $\VSW < -0.8~$V, the switch is opened: no edge channel is transmitted across the switch, and the DCB is fully developed. For $\VSW \in [-0.4,-0.3]~$V, the switch is closed, thus short circuiting the environment: the two outer edge channels are fully transmitted across the switch, and the QPC displays a fixed conductance taken as $\Ginf$. Note that it is necessary to fully open two channels to recover the `intrinsic' conductance: as seen in supplementary Fig.~S1a, the conductance corresponding to one transmitted channel across the switch (intermediate plateau) remains smaller than that measured with two open channels. We do not observe such a difference when a third channel is transmitted across the switch. As discussed in [43], this could be related to the non-negligible charging energy $e^2/2C \sim 40~\mu$eV with respect to the Zeeman splitting $E_\mathrm{Z} \sim 70~\mu$eV. In the widespread `large bias voltage method' (right panel), the DC voltage $V$ across the QPC is increased. We simply assume that at large $V$ the DCB corrections are small and therefore that the conductance converges toward a value taken as $\Ginf$. Note that the two methods generally yield similar results, as shown in Figure~2 of [43].

Now we compare the advantages and drawbacks of both methods. The `large bias voltage method', generally used to infer DCB signal, is also sensitive to possible energy dependences in the `intrinsic' conductance of the quantum coherent conductor (see supplementary note~4). Moreover, heating related to the current flow at large bias voltage and in presence of a large series resistance increases further the voltages for which the `intrinsic' conductance is recovered.
The `switch' method, by just short-circuiting the environment at zero bias voltage avoids these spurious effects. Yet it requires to modify the potential on gates close to the studied QPC: due to capacitive coupling between the gates and the 2DEG, the QPC is slightly modified. The capacitive cross-talk is compensated for by changing simultaneously the voltage $\VQPC$ (applied to the metal split gate controlling the QPC, see article Fig.~1) by an amount proportional to $\VSW$. The proportionality coefficient is calibrated separately: it is adjusted to a fixed value for which the QPC conductance does not change when sweeping $\VSW$ on the switch conductance plateaus at $G_\mathrm{SW}=0$ (resistive environment) and $G_\mathrm{SW}=2/R_\mathrm{q}$ (short-circuited environment). On some samples, such as the one with the series resistance $80$~k$\Omega$, the compensation requires second order corrections and was not perfect. In this $R=80$~k$\Omega$ sample, as well as for the series resistance $R_\mathrm{q}/2$ and $R_\mathrm{q}/3$, we infer $\Ginf$ from the large bias voltage values. In practice, we use the average of the asymptotic conductance extracted at large negative and positive voltages.

\pagebreak
\vspace{\baselineskip}
{\large\noindent\textbf{Supplementary Note 2: Bias voltage dependence in the proposed phenomenological expression for the conductance of a single-channel embedded in a linear environnement}}\\

In the article (article Eq.~1), we have proposed the following phenomenological expression for the differential conductance $G\equiv dI/dV$ of a short single-channel of `intrinsic' conductance $\Ginf$ embedded in a linear environment characterized by the series impedance $Z(\omega)$ (any linear environment can be described as a series impedance and a voltage source as displayed on supplementary Fig.~S6):
\begin{eqnarray}
G(\Ginf,Z,V_\mathrm{DS},T) = \Ginf \frac{1+E_\mathrm{B}(Z,V_\mathrm{DS},T)}{1+R_\mathrm{q}\Ginf E_\mathrm{B}(Z,V_\mathrm{DS},T)},
\label{EqGeneral}
\end{eqnarray}
where $E_\mathrm{B}\equiv\lim_{\Ginf\rightarrow0}\frac{G-\Ginf}{\Ginf}$ is the relative conductance suppression for a single channel in the tunnel regime embedded in the same linear circuit. The quantity $E_\mathrm{B}(Z,V_\mathrm{DS},T)$ can be calculated within the well-known DCB tunnel framework [31].

Note that, strictly speaking, our experimental observation of a relative conductance suppression proportional to $(1-R_\mathrm{q} G)$ at zero DC bias voltage implies the above phenomenological expression at $V_\mathrm{DS}\approx 0$. Because expected heating effects at finite voltage bias limit our experimental accuracy, an ambiguity remains on whether one should inject the generator bias voltage $V_\mathrm{DS}$ or the QPC DC voltage $V$ in the tunnel function $E_\mathrm{B}$ in supplementary Eq.~\ref{EqGeneral} (the conductance remains defined as $G\equiv dI/dV$, and not $ dI/dV_\mathrm{DS}$). Such a distinction is only pertinent beyond the tunnel limit and for series resistances $R$ comparable or higher than $R_\mathrm{q}$ since otherwise $V_\mathrm{DS}\simeq V$.

In the present work, we use $V_\mathrm{DS}$ in supplementary Eq.~\ref{EqGeneral} since with this choice we obtain a perfect agreement with the TLL predictions at $R=R_\mathrm{q}$. In contrast, in the previous work [43] we chose the single-channel DC voltage $V$ and explained by a plausible heating (not taken into account in supplementary Eq.~\ref{EqGeneral}) the deviations observed at intermediate bias voltages for relatively high conductances ($G_\infty \gtrsim 0.5/R_\mathrm{q}$) in series with a large resistance $R=26~\mathrm{k}\Omega$ (see section II.C in the supplementary information of [43]). Note that using the generator bias voltage $V_\mathrm{DS}$ instead of $V$ in supplementary Eq.~\ref{EqGeneral}, as done in the present work, reduces the previously observed deviations between data and the proposed phenomenological expression at intermediate bias voltages (deviations when using $V$ in supplementary Eq.~\ref{EqGeneral}, instead of $V_\mathrm{DS}$, are shown in supplementary Fig.~4 of [43]).

\pagebreak
\vspace{\baselineskip}
{\large\noindent\textbf{Supplementary Note 3: Phenomenological expression for the single-channel conductance using an arbitrary reference point}}\\

The phenomenological expression supplementary Eq.~\ref{EqGeneral} for the single-channel conductance can be straightforwardly recast as a phenomenological scaling law that does not involve necessarily $\Ginf$ but an arbitrary reference point (as first pointed out to us by Y.~Nazarov):
\begin{eqnarray}
\frac{\tau_1/(1-\tau_1)}{\tau_2/(1-\tau_2)}=\frac{1+E_\mathrm{B}(Z,V_\mathrm{DS1},T_{1})}{1+E_\mathrm{B}(Z,V_\mathrm{DS2},T_{2})},
\label{EqScaling}
\end{eqnarray}
where $\tau_{1,2}\equiv G_{1,2}R_\mathrm{q}$ are the conductances, in units of conductance quantum, of the same single-channel quantum conductor in presence of DCB at the different generator bias voltages $V_\mathrm{DS1}$ and $V_\mathrm{DS2}$, and at the different temperatures $T_1$ and $T_2$.

Below, we recast this second formulation (the phenomenological scaling law supplementary Eq.~\ref{EqScaling}) into the expression of the suppressed conductance of a single-channel coherent conductor embedded in a linear environment using an arbitrary reference point (which could be different from the asymptotic $\Ginf$). Knowing the series impedance $Z(\omega)$ of the surrounding linear circuit, the single-channel conductance $\tau_\mathrm{ref}$ at an arbitrary reference point $(V_\mathrm{DSref},T_\mathrm{ref})$ allows one to calculate the full voltage $V_\mathrm{DS}$ and temperature $T$ dependences of the single-channel conductance:
\begin{widetext}
\begin{eqnarray}
\tau (Z,V_\mathrm{DS},T)= \frac{\tau_\mathrm{ref} \left (1+E_\mathrm{B}(Z,V_\mathrm{DS},T) \right )}{1+E_\mathrm{B}(Z,V_\mathrm{DSref},T_\mathrm{ref})+\tau _\mathrm{ref}\times \left (E_\mathrm{B}(Z,V_\mathrm{DS},T)-E_\mathrm{B}(Z,V_\mathrm{DSref},T_\mathrm{ref})\right )}.
\label{EqSI}
\end{eqnarray}
\end{widetext}
Note that the original expression supplementary Eq.~\ref{EqGeneral} corresponds to the limit of a high-energy reference point (at $T,V\rightarrow \infty$, $\tau_\mathrm{ref} \rightarrow R_\mathrm{q}G_\infty$ and $E_\mathrm{B}\rightarrow 0$).

The above formula supplementary Eq.~\ref{EqSI} was used in the article Fig.~4b and article Fig.~5a with a reference point taken at $T=80$~mK in both cases.

\pagebreak
\vspace{\baselineskip}
{\large\noindent\textbf{Supplementary Note 4: `Intrinsic' energy dependences in the conductance across a quantum point contact}}\\

A coherent conductor may present energy dependences in its `intrinsic' conductance $\Ginf$ associated with e.g. a finite dwell time. In the case of QPCs these often result from nearby defects. These energy dependences add up with the DCB energy dependence in presence of the electromagnetic environment, which makes the extraction of the DCB signal as a function of voltage and temperature more difficult. In this note, we illustrate the energy behavior of the QPCs with the electromagnetic environment short-circuited and explain how we deal with the energy dependences of $\Ginf$ in the present work.

The energy dependences of QPCs are usually attributed to defaults in the 2DEG close to the QPC which result in undesired resonances. To minimize such energy dependences, one can tune the magnetic field value within the same quantum Hall plateau or play with the two gate voltages applied to the metal split gates controlling the QPC. Supplementary figure~S2 illustrates the kind of voltage dependences measured in the two extreme representative samples with the series resistance $80$~k$\Omega$ and $R_\mathrm{q}/2$. Note that to access $\Ginf$ in absence of DCB, the two outer edge channels are here fully transmitted across the switch.

For the $R=80~\mathrm{k}\Omega$ sample (supplementary Fig.~S2a), the variations in $R_\mathrm{q}\Ginf$ remain smaller than $0.02$. For the $R_\mathrm{q}/2$ sample (supplementary Fig.~S2b), energy dependencies are rather important at high voltages. When plotting the rescaled curves in the article figure~4a, voltage values where the `intrinsic' conductance was different from the measured conductance at $V=0$ by more than the statistical error bars were not taken into account (open symbols in supplementary Fig.~S2b are excluded for the comparison with the scaling law shown in article Fig.~4a).

Note that temperature dependences of the QPC `intrinsic' conductance also appear at temperature larger than $0.2$~K. This could explain the relatively small discrepancies between data and predictions at higher temperatures in article Fig.~4b.

\pagebreak
\vspace{\baselineskip}
{\large\noindent\textbf{Supplementary Note 5: Heating effects at finite bias}}\\

This note deals with the effect of heating on the suppressed conductance. At low temperature phonons are inefficient to evacuate from the electronic fluid the heat injected locally. Furthermore, the presence of series resistors comparable to the resistance of the QPC impedes the electronic heat currents toward the cold reservoirs (drain, source). Consequently, there are no easy escape paths for the injected Joule power. This results in an increase of the electronic temperature at finite bias voltage that is not taken into account in article Eq.~1 (supplementary Eq.~\ref{EqGeneral}).

A simple approach based on the Wiedemann-Franz law and neglecting dissipation through phonons was developed in the supplementary information of [43] in order to take into account heating. Using this approach, we infer that heating is significant at intermediate bias voltages for the $R=80~\mathrm{k}\Omega$ sample but negligible for a sample with a chromium resistance of value $R_\mathrm{q}/2$ as illustrated on supplementary figure~S3. The dashed lines represent the calculated conductance as a function of the generator bias voltage $V_\mathrm{DS}$ taking into account heating in the chromium resistance for a transmission $\tauinf=0.23$, a base temperature of $25~$mK, a parallel capacitance of $2~$fF and the two resistances values. The continuous lines are the conductances calculated using article Eq.~1 (supplementary Eq.~\ref{EqGeneral}). As a consequence of such heating, the data of the $R=80~\mathrm{k}\Omega$ sample do not obey the phenomenological scaling law supplementary Eq.~\ref{EqScaling} at intermediate bias voltages, contrary to the other samples.

\pagebreak
\vspace{\baselineskip}
{\large\noindent\textbf{Supplementary Note 6: Supplementary comparison with results obtained on a carbon nanotube resonant level embedded in a resistive environment [44, 45]}}\\

In this note, we further compare and discuss the predictions of the scaling law article Eq.~2 (supplementary Eq.~\ref{EqScaling}) with the results presented for the conductance of a carbon nanotube embedded in a dissipative environment measured in [44] and [45].

\vspace{\baselineskip}
{\noindent\textbf{Equilibrium data ($V\approx 0$) [44].}}\\

In article Fig.~5a, we compared the data obtained in [44] by changing the energy of the carbon nanotube single-channel discrete level (data shown in Fig.~4a of [44]). On supplementary Fig.~S4a, we perform the same comparison but with the data of [44] obtained by tuning the symmetry of the coupling between the nanotube and the leads (data shown in Fig.~3a of [44]). Importantly, as in the article, the predictions derived from the phenomenological expression supplementary Eq.~\ref{EqSI} (continuous lines) are computed quantitatively essentially without any fit parameters. The calculations are performed using the series resistance $R=0.75R_\mathrm{q}$ [44], assuming a small parallel capacitance $C=0.07~\mathrm{fF}$ (there is little effect for realistic values $C\lesssim0.1~\mathrm{fF}$) and using the conductance measured at $T_\mathrm{ref}=80~\mathrm{mK}$ as the reference point in supplementary Eq.~\ref{EqSI}.

We find a reasonable agreement between data and phenomenological expression supplementary Eq.~\ref{EqSI} at low enough temperatures such that the symmetric resonant level ($\blacksquare$) is perfectly transmitted. Note that the conductance reduction that appears in the symmetric case at higher temperatures signals a transition toward either the sequential tunneling regime [44] or the long dwell-time regime with energy dependent transmissions [45], where the phenomenological expression supplementary Eq.~\ref{EqSI} for short quantum conductors does not hold.

On supplementary figure~S4b, to further test the proposed phenomenological expression supplementary Eq.~\ref{EqSI}, we confront in the case of a symmetric coupling to the dissipative leads the resonant lineshapes at different temperatures (measured as a function of the plunger gate voltage controlling the energy of the discrete level, data of Figure~2b in [44] shown as symbols in supplementary Fig.~S4b) with the predictions of supplementary Eq.~\ref{EqSI} (lines). The quantitative predictions of supplementary Eq.~\ref{EqSI} are obtained using the characterized environment ($R=0.75R_\mathrm{q}$ and $C=0.07~\mathrm{fF}$) and using as reference points the data at $T_\mathrm{ref}=0.4$~K.

\vspace{\baselineskip}
{\noindent\textbf{Out-of-Equilibrium data [45].}}\\

In [45], non-equilibrium effects are investigated on the same system by applying a voltage bias to the nanotube. The proposed phenomenological expression supplementary Eq.~\ref{EqGeneral} also accounts for non-equilibrium conductances. On supplementary Fig.~S5, we compare the data of Fig.~S1 in [45] with the prediction of supplementary Eq.~\ref{EqSI}.

The calculations (lines in supplementary Fig.~S5) are performed using the series resistance $R=0.75 R_\mathrm{q}$ [44], assuming a small parallel capacitance $C=0.07~\mathrm{fF}$, and using as a reference point in supplementary Eq.~\ref{EqSI} the conductance measured at zero voltage bias ($V_\mathrm{DSref}=0$). The temperatures are set to $T=50~\mathrm{mK}$ for the left panel displaying data obtained for different coupling asymmetries, and to $T=45~\mathrm{mK}$ for the right panel displaying data obtained with a symmetric coupling but different energies of the discrete level [45]. Importantly, the predictions derived from the phenomenological scaling law are computed with the temperature as the only fit parameter. The value $50~\mathrm{mK}$ corresponds to the indicated temperature in the article [45] but the data in the symmetrically coupled case (right panel), being sharper, require the slightly smaller temperature $T=45~\mathrm{mK}$ to be accounted for. We find a reasonable agreement between data and the phenomenological expression supplementary Eq.~\ref{EqSI}. Note that the conductance decrease observed above $20~\mu V$ close to $G\simeq 1/R_\mathrm{q}$ signals a transition toward either the sequential tunneling regime [44] or the long dwell time regime with energy dependent transmissions [45], where the phenomenological expression supplementary Eq.~\ref{EqScaling} for short quantum conductors does not hold.

\vspace{\baselineskip}
{\noindent\textbf{Comparison with the renormalization group result of \cite{aristov2008, aristov2009}, which was found consistent with the equilibrium scaling law observed by Mebrahtu \emph{et al.} [45].}}\\

Mebrahtu and coauthors observed in [45] that, below 400~mK, the temperature dependence of the conductance data obtained when tuning the discrete level out-of-resonance could be recast as a function of the single parameter $\Delta V_\mathrm{gate}/T^{R/(R+R_\mathrm{q})}$, where $\Delta V_\mathrm{gate}$ is the variation of the plunger gate voltage controlling the energy of the discrete level (see Fig.~4a in [45]). They obtained a good fit of the low temperature data (see Fig.~4a in [45]) assuming the conductance obeys the following expression (Eq.~2 in [45]):
\begin{equation}
\frac{T}{(\Delta V_\mathrm{gate})^{1+R_\mathrm{q}/R}} \propto \frac{\tau^\frac{R_\mathrm{q}}{2R}}{(1-\tau)^\frac{R+R_\mathrm{q}}{2R}}\sqrt{\frac{1+\tau R/R_\mathrm{q}}{1+R/R_\mathrm{q}}}.
\label{scalingT_Mebrahtu}
\end{equation}

The above expression is inspired by the renormalization group result obtained by \cite{aristov2008, aristov2009} within the TLL framework. The approximate result of Aristov and W\"olfe was derived for an arbitrary Luttinger interaction coefficient $K=1/(1+R/R_\mathrm{q})$, corresponding to purely dissipative environnements characterized by a series resistance R of arbitrary value [19], and for arbitrary renormalized transmissions $\tau(T)$. It reads (Eq.~66 in \cite{aristov2009}):
\begin{equation}
\frac{T}{T_\mathrm{ref}}=\left(\frac{\tau}{\tau_\mathrm{ref}}\right)^\frac{R_\mathrm{q}}{2R}\left(\frac{1-\tau_\mathrm{ref}}{1-\tau}\right)^\frac{R+R_\mathrm{q}}{2R}\left(\frac{1+\tau R/R_\mathrm{q}}{1+\tau_\mathrm{ref} R/R_\mathrm{q}}\right)^\frac{4c_3}{2},
\label{scalingT_Aristov}
\end{equation}
with $c_3$ a non universal parameter depending on cutoff choices, and $\tau_\mathrm{ref}(T_\mathrm{ref})$ the initial condition. According to Aristov and W\"olfe \cite{aristov2008, aristov2009}, $c_3=1/4$ provides the best fit of the exact thermodynamic Bethe ansatz solution at $R=R_\mathrm{q}$.

We now directly compare the renormalization group result of Aristov and W\"olfe (supplementary Eq.~\ref{scalingT_Aristov}) with the proposed phenomenological scaling law supplementary Eq.~\ref{EqScaling}. Assuming a purely dissipative environnement $Z(\omega)=R$, zero DC bias and the reference point $\tau_\mathrm{ref}(T_\mathrm{ref})$, supplementary Eq.~\ref{EqScaling} reads:
\begin{equation}
\frac{T}{T_\mathrm{ref}}=\left(\frac{\tau}{\tau_\mathrm{ref}}\frac{1-\tau_\mathrm{ref}}{1-\tau}\right)^\frac{R_\mathrm{q}}{2R}.
\label{scalingT_us}
\end{equation}

First note that these scaling laws (supplementary Eqs~\ref{scalingT_Aristov} and \ref{scalingT_us}) match for $R\ll R_\mathrm{q}$ at arbitrary values of $\tau$ and $\tau_\mathrm{ref}$, and also for $\{\tau , \tau_\mathrm{ref}\}\ll 1$ at arbitrary values of $R$. In general, however, there are deviations. In particular at $R=R_\mathrm{q}$, for which the parameter $c_3=1/4$ was adjusted to fit the exact thermodynamic Bethe ansatz solution \cite{aristov2008, aristov2009}, and where the proposed phenomenological scaling law supplementary Eq.~\ref{EqScaling} corresponds exactly to the thermodynamic Bethe ansatz solution, the two scaling laws still deviate by the factor
$\sqrt{\frac{1+\tau}{1-\tau}\frac{1-\tau_\mathrm{ref}}{1+\tau_\mathrm{ref}}}$. Note that, in most cases, this discrepancy in the scaling laws at $R=R_\mathrm{q}$ does not result in large absolute differences on the calculated values of $\tau$.

\pagebreak
\vspace{\baselineskip}
{\large\noindent\textbf{Supplementary Note 7: Generalization of the mapping between a single channel in series with a resistance and the problem of an impurity in a TLL}}\\

The considered situation is displayed in the schematic circuit supplementary Fig.~S6.

The action describing quantum transport across a short single-channel quantum conductor can be written in terms of a single local bosonic field denoted $\hat Q(t)$ [19]. This bosonic field $\hat Q(t)$ can be identified with the total charge transferred across the conductor and therefore it is directly related to the current operator $\hat I(t)=\partial_t \hat Q(t)$. For simplicity we focus here on the zero-temperature limit.

First in absence of the environment, the action describing transport across the single-channel conductor is the sum of the two contributions of the leads ($\mathcal S_0$) and the single-channel conductor ($\mathcal S_\mathrm{B}$) $\mathcal S =\mathcal S_0+\mathcal S_\mathrm{B}$. Within a linear spectrum approximation, the leads contribution reads:
\begin{eqnarray}\label{szero}
  \mathcal{S}_0 = \frac{\hbar}{e^2}\int d\omega |\omega||\hat Q(\omega)|^2\,.
\end{eqnarray}
Regarding the single-channel conductor action, following Kane and Fisher (Eq.~3.2 in \cite{kane1992prb}) (see also Fendley, Ludwig and Saleur, Eq.~3.4 in [47]) a potential scatterer $v(x)$ of arbitrary strength coupled to the electronic density and which is nonzero only for x near zero, is described by the action:
\begin{eqnarray}\label{sbgeneral}
\mathcal{S}_\mathrm{B} = \sum_{n=-\infty}^{\infty} \int dt \, v_{\mathrm{B}n} \, e^{2 i \pi n \hat Q(t)/ e}\, ,
\end{eqnarray}
where the coefficients $v_{\mathrm{B}n}$ are proportional to the Fourier transform of $v(x)$ at momenta given by $n2k_\mathrm{F}$, with $k_\mathrm{F}$ the fermi wave vector.
It is a usual practice to neglect the terms at $|n|\geq 2$ and use the following simplified single-channel conductor action (see e.g. [19,~47], and also the supplementary references \cite{kane1992prb,egger1995,weiss1996}):
\begin{eqnarray}\label{sb}
  \mathcal{S}_\mathrm{B} = \int dt \, v_\mathrm{B} \, \cos (2 \pi \hat Q(t)/ e)\,,
\end{eqnarray}
where we have introduced the backscattering amplitude $v_\mathrm{B}$. Note first that the above simplified single-channel conductor action is sufficient to emulate the full range of intrinsic transmission probabilities $\tau_\infty \in [0,1]$, from $\tau_\infty \rightarrow 0$ at $v_\mathrm{B} \rightarrow \infty$ to $\tau_\infty \rightarrow 1$ at $v_\mathrm{B} \rightarrow 0$ (see e.g. \cite{egger1995} for a specific discussion regarding this point\footnote{This point can be further strengthen by two remarks: 1) a mapping toward the same TLL problem, with the same interaction parameter, can be obtained by describing the single-channel quantum conductor with a tunnel Hamiltonian weakly connecting two semi-infinite 1D wires [19]. 2) It can be shown using a renormalization group approach \cite{kane1992prb} that only the terms $n^2<(1+R/R_\mathrm{q})$ in the full single-channel conductor action (supplementary Eq.~\ref{sbgeneral}) are relevant at low energy for a Luttinger interaction coefficient $1/(1+R/R_\mathrm{q})$. Consequently for $R<3R_\mathrm{q}$ and at low energy the supplementary Eqs~\ref{sbgeneral} and \ref{sb} are effectively equivalent. This last statement is stronger than saying that the simplified action supplementary Eq.~\ref{sb} covers the full range of backscattering strengths. In general, supplementary Eqs~\ref{sbgeneral} and \ref{sb} are not exactly equivalent.}).
Note also that bosonization is justified only below the energy cutoff $\hbar \omega_\mathrm{F}$ that delimits the validity domain for both the short single-channel conductor approximation (limited by the finite dwell time across the conductor and the energy barrier separating additional electronic channels) and the linearization of the energy spectrum in the leads. For instance, with edge states in the integer quantum Hall regime, $\omega_\mathrm{F}$ is limited at least by the cyclotron frequency.

We now introduce the coupling between the conductor and a series impedance: $\hat Q(t) (V_\mathrm{DS}- \hat u(t))$, where $\hat u(t)$ is the voltage drop across the impedance, and $V_\mathrm{DS}$ the voltage imposed by the generator. This corresponds to the coupling action:
\begin{equation}
\mathcal{S}_\mathrm{c}=\int dt \hat Q(t) (V_\mathrm{DS}-\hat u(t)).
\end{equation}

Finally, the time dependence of $\hat u(t)$ is governed by the environmental gaussian action $ \mathcal{S}_\mathrm{\hat u}$. Using the fluctuation-dissipation theorem, $\int_{-\infty}^{\infty}e^{i\omega t} \left < \hat u(t) \hat u(0)+\hat u(0) \hat u(t)\right >\,dt=2 \hbar | \omega | \mathrm{Re}[Z(\omega)]$, one obtains \cite{leggett1984,falci1991}:
\begin{equation}
\mathcal{S}_\mathrm{\hat u}=\int \frac{d\omega}{(2\pi)^2} \frac{|\hat u(\omega)|^2}{|\omega| \mathrm{Re}[Z(\omega)]}.
\end{equation}

Thus the total action reads $\mathcal{S}=\mathcal{S}_0+\mathcal{S}_\mathrm{B}+\mathcal{S}_\mathrm{\hat u}+\mathcal{S}_\mathrm{c}$. By integrating out $\hat u(t)$, one obtains an effective action for $\hat Q(t)$  describing quantum transport across an arbitrary short single-channel conductor in series with any linear electromagnetic environment characterized by an impedance $Z(\omega)$:
\begin{equation}\label{envbis}\begin{split}
\mathcal{S}_\mathrm{eff}=&\frac{\hbar}{e^2}\int d\omega |\omega| \left(1+\frac{\mathrm{Re}[Z(\omega)]}{R_\mathrm{q}}\right)|\hat Q(\omega)|^2\\
&+\mathcal{S}_\mathrm{B}+\int dt\, \hat Q(t)V_\mathrm{DS}.
\end{split}
\end{equation}

In the limit of a purely dissipative environnement $Z(\omega)=R$, this action corresponds precisely to the conventional TLL model with the Luttinger interaction coefficient $K=1/(1+R/R_\mathrm{q})$.

Remarkably, a frequency dependent circuit impedance $Z(\omega)$ corresponds to the more general problem of a 1D conductor with finite range electron-electron interactions $V(x)$. As shown in \cite{fabrizio1994}, a finite range electron-electron interaction can be encapsulated into the action as a frequency dependent effective Luttinger interaction coefficient $K(\omega)$ (see Eqs~6 and 7 in \cite{fabrizio1994}). The exact same action is here derived for the corresponding circuit impedance given by $K(\omega)=1/(1+ \mathrm{Re}[Z(\omega)]/R_\mathrm{q})$. More specifically, we find along the lines of Giamarchi [2] that a circuit impedance $Z(\omega)$ corresponds to a finite-range impurity potential that obeys (see Eqs~4.4, 10.44, and 10.46  in [2]):
\begin{equation}
\label{finiterange}
\frac{1}{1+\frac{\mathrm{Re}[Z(\omega)]}{R_\mathrm{q}}}=\int dq \frac{v_\mathrm{F} |\omega|/\pi}{\omega^2+q^2 v_\mathrm{F}^2 (1+2V(q)/\pi \hbar v_\mathrm{F})},
\end{equation}
with $V(q)$ the spatial Fourier transform of the interaction potential $V(x)$ and $v_\mathrm{F}$ the Fermi velocity.

We now assume that the environmental impedance can be expanded into power series of $\omega$:
\begin{equation}
\label{expansion}
\mathrm{Re}[Z(\omega)]=R+\sum_{n=1}^\infty R_n\left(\omega/\omega_\mathrm{Z}\right)^n,
\end{equation}
where $\mathrm{Re}[Z(\omega=0)]=R$ and $\omega_\mathrm{Z}$ is the cutoff frequency (the convergence radius) of the Taylor series expansion, which coincides with $1/RC$ for a $RC$ circuit.
Then we make the following crucial statement: by power counting, one can argue that the higher order terms of the expansion, once injected into $S_\mathrm{eff}$ (supplementary Eq.~\ref{envbis}) yield less relevant contributions compared to the term associated to $R/R_\mathrm{q}+1$. Thus one can replace $\mathrm{Re}[Z(\omega)]$ by $R$ in supplementary Eq.~\ref{envbis}, yielding a similar action to that for the impurity  problem in a TLL with an interaction parameter $1/(1+R/R_\mathrm{q})$, provided one works at frequency scales below $\min \{\omega_\mathrm{F},\omega_\mathrm{Z}\}$.

We therefore established that the mapping remains valid in more realistic situations than purely dissipative environment, in presence of a high frequency cut-off in the environmental impedance.

\pagebreak
\vspace{\baselineskip}
{\large\noindent\textbf{Supplementary Note 8: Exact solution for the out-of-equilibrium current compared to the flow equation article Eq.~3}}\\

The mapping to a TLL allows to exploit the thermodynamic Bethe ansatz exact solution (see [46,~47] and the supplementary reference \cite{weiss1996}).
We focus here on the out-of-equilibrium regime corresponding to $k_\mathrm{B} T\ll e V_\mathrm{DS}$ (including finite temperature results in an involved integro-differential equation that can be solved numerically but not analytically).
Following Fendley \textit{et al} [46,~47], all the dependence on the backscattering amplitude ($v_\mathrm{B}$ in supplementary Eq.~\ref{sb}) is encoded into the voltage scaling parameter $V_\mathrm{B}$ ($V_\mathrm{B}$ in the present work corresponds to $2k_\mathrm{B}T_\mathrm{B}/e$ in [46,~47])\footnote{In general, there is no universal relation between $V_\mathrm{B}$ and the `intrinsic' transmission probability $\tau_{\infty}$. However, a useful approximate expression is $V_\mathrm{B} \propto (1/\tau_\infty-1)^{R_\mathrm{q}/2R}$.}. The average current $I=<\hat I>$ is written in terms of two power series that are different for the two regimes of high and low energies:
\begin{equation}\label{eq-current}
  I(V_\mathrm{DS})=\left\{
  \begin{array}{l}
    \displaystyle \frac{e^2\, V_\mathrm{DS}}{h(1+r)}\left[1-\frac{1}{1+r}\sum_{n=1}^\infty a_n\left (\frac{1}{1+r}\right )\left(\frac{eV_\mathrm{DS}}{k_\mathrm{B} T_\mathrm{B}^{'}}\right)^{\frac{-2nr}{r+1}}\right] \\ \\
    \displaystyle \frac{e^2\, V_\mathrm{DS}}{h}\sum_{n=1}^\infty a_n(1+r)\left(\frac{eV_\mathrm{DS}}{k_\mathrm{B}T_\mathrm{B}^{'}}\right)^{2nr}\; ,
  \end{array}
  \right.
\end{equation}
where the first (second) line corresponds to $eV_\mathrm{DS}$ larger (smaller) than the crossover energy $k_\mathrm{B} T_\mathrm{B}^{'}\,e^\Delta$ at which both series match. Here  $r=R/R_\mathrm{q}$, $T_\mathrm{B}^{'}$ is related to the widely used characteristic voltage $V_\mathrm{B}$ via the following equation
\begin{equation}
  T_\mathrm{B}^{'}=\frac{eV_\mathrm{B}}{k_\mathrm{B}} \,\frac{\sqrt{\pi}(r+1)\Gamma(\frac{1}{2}+\frac{1}{2r})}{\Gamma(\frac{1}{2r})}\,,
\end{equation}
and
\begin{equation}
  \Delta=\frac{1}{2}\log(r)-\frac{1+r}{2r}\log(1+r)\,.
\end{equation}
The functions $a_n(x)$ in supplementary Eq.~\ref{eq-current} are given by:
\begin{equation}
  a_n(x)=(-1)^{n+1}\frac{\sqrt{\pi}\,\Gamma(n x)}{2\Gamma(n)\Gamma(\frac{3}{2}+n(x-1))}\,.
\end{equation}

Note that the coherent conductor differential conductance $G=dI/dV$, with $V$ the DC voltage across the single-channel conductor, can be readily derived from $I(V_\mathrm{DS})$, with $V_\mathrm{DS}$ the generator bias voltage: using $V(V_\mathrm{DS})=V_\mathrm{DS}-RI(V_\mathrm{DS})$, one obtains $G(V_\mathrm{DS})=G_\mathrm{tot}(V_\mathrm{DS})/(1-RG_\mathrm{tot}(V_\mathrm{DS}))$ with $G_\mathrm{tot}(V_\mathrm{DS})=dI(V_\mathrm{DS})/dV_\mathrm{DS}$ the total conductance of the circuit.

Note also that the differential conductances $G$ and $G_\mathrm{tot}$ only depend on the scaled voltage $V_\mathrm{DS}/V_\mathrm{B}$ and on the parameter $r=R/R_\mathrm{q}$: the same universal function therefore covers the full range of `intrinsic' transmission probability $\tau_\infty \in [0,1]$ by varying the scaling parameter $V_\mathrm{B}$ from $\infty$ to 0. Interestingly, a similar conclusion can be reached using a renormalization group analysis (see e.g. section~IV in [20]).

The current-voltage relationship in the two limits of $eV_\mathrm{DS}\gg k_\mathrm{B} T_\mathrm{B}^{'}$ and $eV_\mathrm{DS}\ll k_\mathrm{B} T_\mathrm{B}^{'}$ are obtained by considering up to first order the corresponding power series in supplementary Eq.~\ref{eq-current}  \cite{kane1992}.
In the regime of suppressed transmission ($eV_\mathrm{DS}\ll k_\mathrm{B} T_\mathrm{B}^{'}$), the current reads [19]:
\begin{equation}\label{tunneling}
I(V_\mathrm{DS})=\frac{e^2\, V_\mathrm{DS}}{h} a_1(1+r)\left(\frac{eV_\mathrm{DS}}{k_\mathrm{B}T_\mathrm{B}^{'}}\right)^{2r}.
\end{equation}
Remarkably, this corresponds to the power law predicted for tunnel junctions within the $P(E)$ theory [31], in agreement with the phenomenological expression article Eq.~1. One can check directly that it verifies the flow equation article Eq.~3 and the phenomenological scaling law for a series resistance $R$ article Eq.~4.

Next, we consider the specific value $R=R_\mathrm{q}$ at which the series in supplementary Eq.~\ref{eq-current} can be summed up to obtain the analytical expression :
\begin{equation}\label{transmission_r_1}
 I(V_\mathrm{DS})=\frac{V_\mathrm{B}}{2R_\mathrm{q}}\left [\frac{V_\mathrm{DS}}{V_\mathrm{B}}-\arctan \left (\frac{V_\mathrm{DS}}{V_\mathrm{B}}\right ) \right ],
\end{equation}
which obeys exactly the flow equation article Eq.~3 and the phenomenological scaling law for a series resistance $R$ article Eq.~4.

\pagebreak
\vspace{\baselineskip}
{\large\noindent\textbf{Supplementary Note 9: Relationship between conductance suppression and quantum shot noise}}\\

Insight can be gained on the flow equation article Eq.~3 by recalling a fundamental link established between the DCB and shot-noise.
Qualitatively, the presence of shot-noise appears as a necessary ingredient to excite the electromagnetic modes of the circuit, and therefore to the very existence of DCB.
In its initial form, such a direct link was derived theoretically in the perturbative limit of small series impedances [32,~33]: the reduction of the conductance due to DCB was shown to be proportional to the intrinsic shot-noise of the conductor (i.e. the shot noise without the environment). Nevertheless, such a perturbative relation led to a logarithmic divergence at low voltages.
The mapping to a TLL has allowed to propose a generalized and exact non-perturbative link for arbitrary values of $R$, obeyed at all voltages within the domain of validity of the mapping [19]. Interestingly, in the particular case $R=R_\mathrm{q}$ where the problem can be recast in that of free fermions with an energy dependent transmission $\tau (E)$, using a procedure called refermionization (see e.g. [53-55] and the supplementary references \cite{furusaki1997,vondelft1998,egger1998,aristov2009}), the above mentioned generalized link with shot noise follows directly from the usual  reduction factor (Fano factor) $\tau(E)(1-\tau(E))$ of the Poisson noise. Note that the thermodynamic Bethe ansatz solution for the noise at finite temperatures and voltages has previously been compared to a similar scattering approach approximate expression \cite{trauzettel2004}. At arbitrary values of $R$, one exploits the thermodynamic Bethe ansatz solution for the out-of-equilibrium zero-frequency noise [46], which can be expressed only in terms of the average current and the circuit differential conductance $G_\mathrm{tot}$:
\begin{equation}
  \frac{d\, G_\mathrm{tot}(V_\mathrm{DS})}{d\log V_\mathrm{DS}}=\frac{2 R}{e R_\mathrm{q}} \,\frac{d S(V_\mathrm{DS})}{dV_\mathrm{DS}}\,,
  \label{eq_scal_exact}
\end{equation}
with $S(V_\mathrm{DS})=\int dt\,[\langle I(0)I(t)\rangle-\langle I(t)\rangle^2]$.
We stress here that the differential shot-noise is that in the presence of the environment, and depends on both $R$ and $V_\mathrm{DS}$. This relation is one example of a more general connection (see [19] and the supplementary reference \cite{saleur2001}) between higher order cumulants of the current.
Thus even for a low resistance $R$, supplementary Eq.~\ref{eq_scal_exact} offers a non-perturbative link between DCB and noise, and describes the strong back-action of the environment.

\clearpage
\section{Supplementary References}